\documentclass[11pt]{elsarticle}

\usepackage{graphicx} 
\usepackage{ amsthm, amsmath, amsfonts, bbm, enumerate, indentfirst}
\usepackage{caption, subcaption}
\graphicspath{ {images/} }
\usepackage{color} 
\usepackage{multirow}

\usepackage{tikz}
\usetikzlibrary{arrows}

\usepackage{ algpseudocode, algorithm}
\usepackage{doi}
\usepackage{hyperref}
\newcommand\mX{\ensuremath{\mathfrak{X}}}
\newcommand\mx{\ensuremath{\mathbf{x}}}
\newcommand\mS{\ensuremath{\mathfrak{S}}}
\newcommand\cS{\ensuremath{\mathcal{S}}}
\newcommand\cX{\ensuremath{\mathcal{X}}}
\newcommand\cF{\ensuremath{\mathcal{F}}}
\newcommand\cG{\ensuremath{\mathcal{G}}}
\newcommand\cZ{\ensuremath{\mathcal{Z}}}
\newcommand\hmS{\hat{\mathfrak{S}}}

\DeclareMathOperator\dimn{dim}

\newtheorem{remark}{Remark}

\makeatletter
\def\@author#1{\g@addto@macro\elsauthors{\normalsize%
    \def\baselinestretch{1}%
    \upshape\authorsep#1\unskip\textsuperscript{%
      \ifx\@fnmark\@empty\else\unskip\sep\@fnmark\let\sep=,\fi
      \ifx\@corref\@empty\else\unskip\sep\@corref\let\sep=,\fi
      }%
    \def\authorsep{\unskip,\space}%
    \global\let\@fnmark\@empty
    \global\let\@corref\@empty  
    \global\let\sep\@empty}%
    \@eadauthor={#1}
}
\makeatother

\begin{document}
\newcommand{\indicator}[1]{\mathbbm{1}_{\left\{{#1} \right\}}}

\begin{frontmatter}
\title{Bayesian Epidemic Detection in Multiple Populations}

\author{Katherine Shatskikh\corref{cor1}}
\ead{shatskikh@pstat.ucsb.edu}
\cortext[cor1]{Corresponding author}

\author{Michael Ludkovski}
\ead{ludkovski@pstat.ucsb.edu}

\address{Department of Statistics and Applied Probability, University of California Santa Barbara, \\ Santa Barbara, CA 93106-3110, U.S.A.}

\begin{abstract}
Traditional  epidemic detection algorithms make decisions using only local information. We propose a novel approach that explicitly models spatial information fusion from several metapopulations. Our method also takes into account cost-benefit considerations regarding the announcement of epidemic. We utilize a compartmental stochastic model within a Bayesian detection framework which leads to a dynamic optimization problem. The resulting adaptive, non-parametric detection strategy optimally balances detection delay vis-a-vis probability of false alarms. Taking advantage of the underlying state-space structure, we represent the stopping rule in terms of a detection map
which visualizes the relationship between the multivariate system state and policy making. It also allows us to obtain an efficient simulation-based solution algorithm that is based on the Sequential Regression Monte Carlo (SRMC) approach of Gramacy and Ludkovski (SIFIN, 2015). We illustrate our results on synthetic examples and also quantify the advantages of our adaptive detection relative to conventional threshold-based strategies.
\end{abstract}

\begin{keyword} Biosurveillance; quickest detection; regression Monte Carlo; stochastic compartmental models;
\end{keyword}
\end{frontmatter}

\section{Introduction}
\label{section: intro}

Infectious disease epidemics intrinsically unfold across both space and time. As a result, biosurveillance algorithms need to integrate spatio-temporal data. This is especially so in the context of statistical inference, whereby syndromic surveillance at neighboring locales carries additional information that can be fused for improved decision making in terms of initiating and organizing epidemic counter-measures. A crucial first step for response strategies is to identify,  or detect, in real-time the epidemic outset. In this article, we propose a methodology that allows for such \emph{optimal} decision-making with spatial information fusion.
Specifically, we investigate a model that combines quickest detection with a spatial metapopulation setup, integrating information received from multiple geographic domains. To reflect the inherent uncertainty in epidemic evolution (which is amplified under partial information), we develop a stochastic compartmental (or state-space) epidemic model, which allows us to generate adaptive, nonparametric detection rules. Extant approaches largely propose heuristic detection strategies, concentrating primarily on the inferential aspect of the statistical model \cite{Chan2010,lin2013sequential,Niemi2014,Skvortsov2012}. For instance, a typical approach is to announce an epidemic as soon as the estimated number of infecteds in the local population is above a fixed $\bar{I}$. In contrast, we dynamically optimize the detection strategy,  to come up with a ``best'' detection rule within our mechanistic outbreak model.

Traditional compartmental epidemic models deal with a single population; the spatial aspect is treated by building a series of such single-population models that are estimated/forecasted independently. This is also a common surveillance approach, especially for recurring infectious epidemics, such as influenza-like illness (ILI), dengue fever, or measles. For example, in the US the existing biosurveillance systems for flu operate primarily at the state level and are siloed across states. This limitation of existing practice was brought into sharp relief during the 2014 Ebola outbreak in West Africa. The epidemic has been accompanied by a dearth of reliable information, leading to extreme spread in forecasts regarding the future course of the outbreak. In addition, numerous statistical methods \cite{WHOEbola, chowell2014transmission, fisman2014early} were put forth attempting to infer in ``real-time'' the actual size and parameters of the outbreak in different locales. However, nearly all these methods were single-population, so that when trying for example to infer the number of Ebola infecteds in Liberia, only Liberian data was utilized, completely ignoring similar and highly relevant data from neighboring Guinea and Sierra Leone. Similarly, at the more granular provincial level, data from neighboring provinces was generally not used during estimation procedures.

\begin{figure}[ht] \hspace*{-10pt}
\begin{tabular}{lcr}
\includegraphics[width=0.3\textwidth]{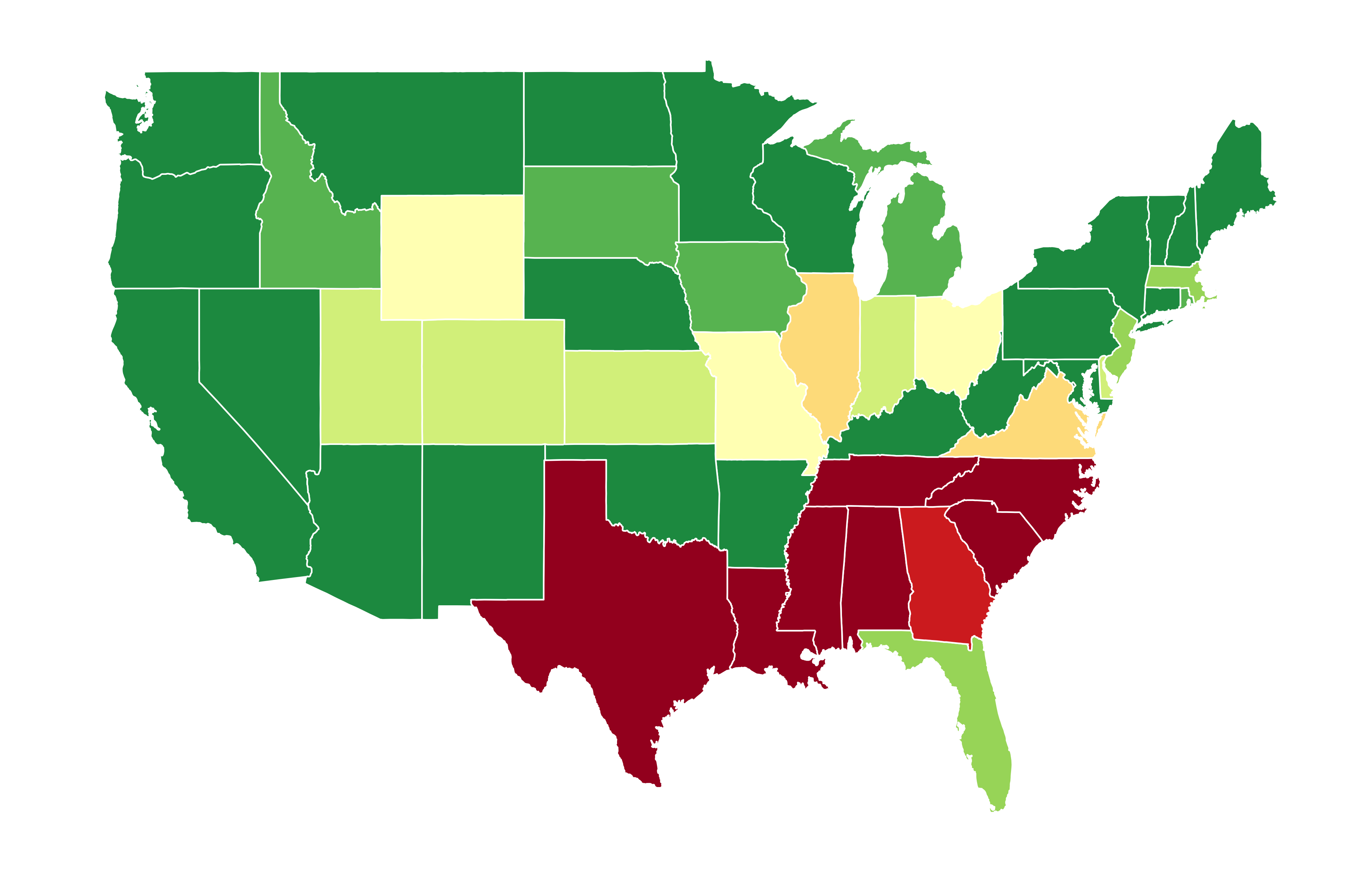} &
\includegraphics[width=0.3\textwidth]{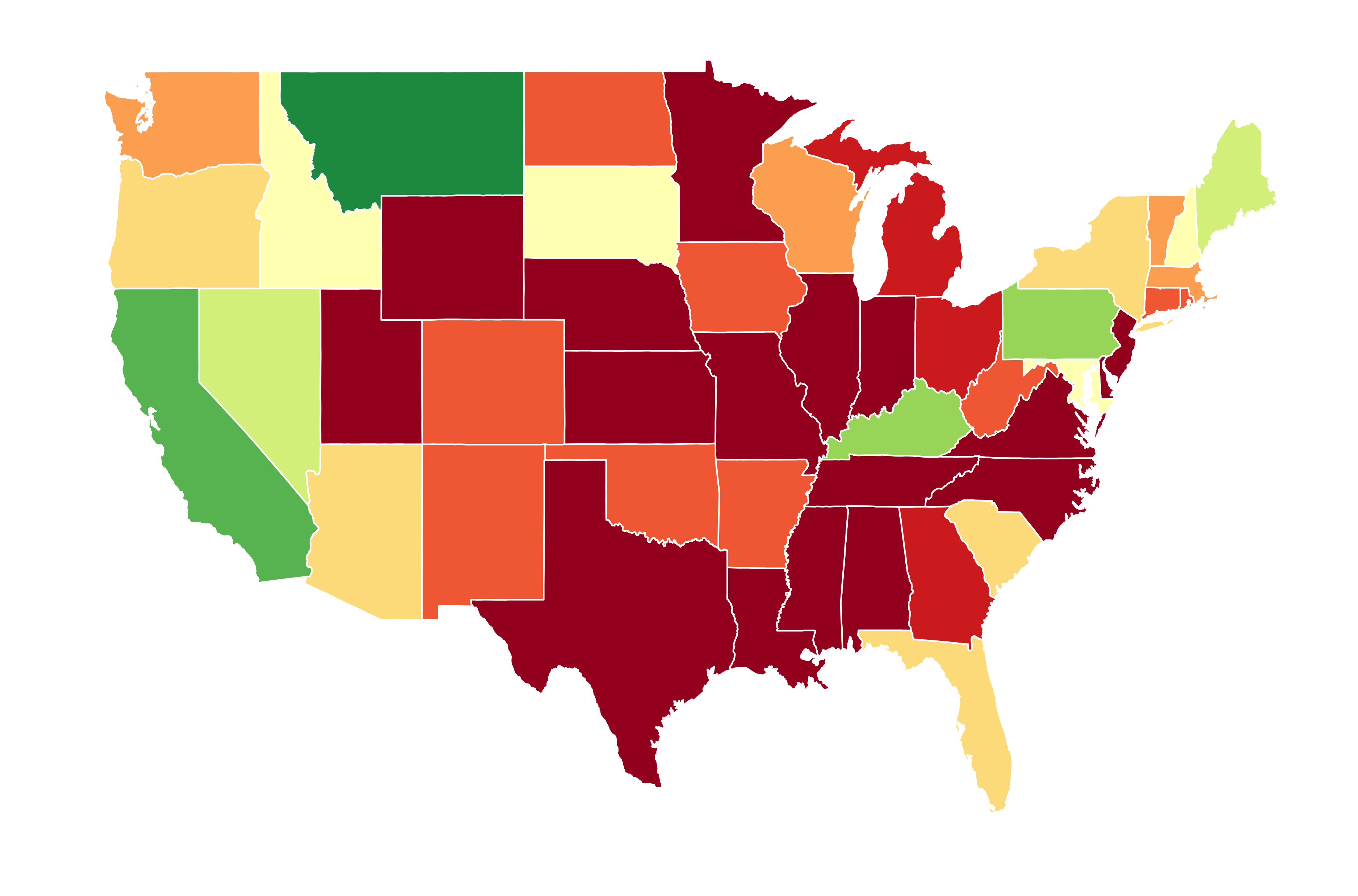} &
\includegraphics[width=0.3\textwidth]{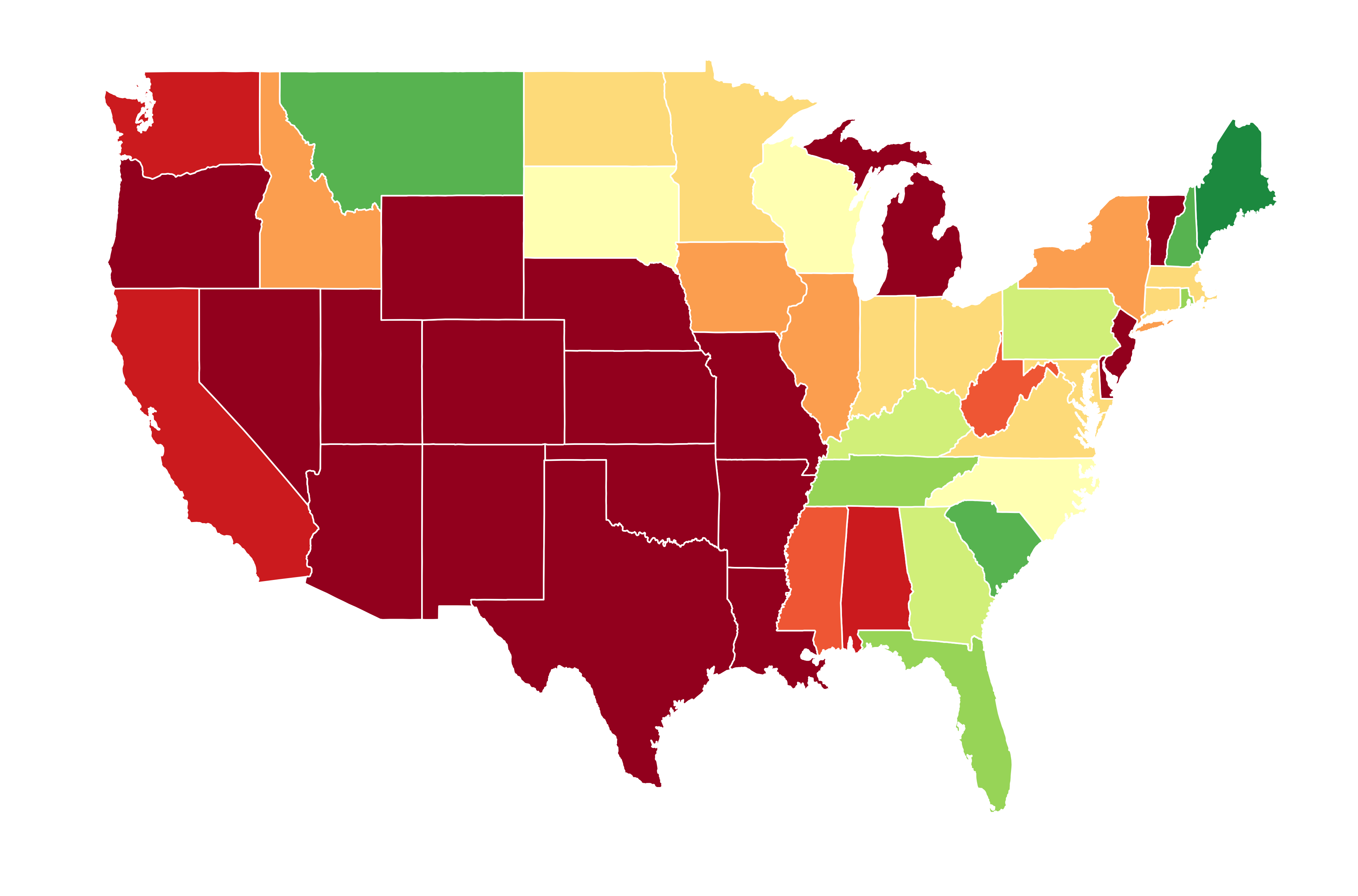}\\
Week 49, 2012 & Week 1, 2013 & Week 4, 2013
\end{tabular}
\caption{Spread of Influenza during the 2012-13 Flu season according to FluView CDC data. The colors represent weekly ILI activity levels in terms of percentage of doctor visits attributed to ILI relative to low-season baseline. Green indicates at/below mean, while shades of red indicate outbreak activity (with darkest color corresponding to eight or more standard deviations above the mean). Weeks are numbered from January 1, and are 12/3-9/2012 (Week 49), 12/31/2012-1/6/2013 (Week 1) and 1/21-27/2013 (Week 4), respectively. Data source: \url{http://www.cdc.gov/flu/weekly/pastreports.htm}.
\label{fig:epidemic-movie}}
\end{figure}

For a less dramatic and perhaps more statistically convenient example, we discuss the yearly influenza outbreaks in United States. Figure \ref{fig:epidemic-movie} illustrates the spatial dynamics of ILI during the 2012-13 flu season. As can be observed, the peak of the outbreak varied significantly (up to 6-8 weeks difference) across different parts of the country. Nevertheless, there is a clear propagation, making spatial information fusion desirable. Figure \ref{fig:epidemic-movie} indicates that the current, single-population based detection protocols are not sufficient; for instance the fact that there are increased ILI levels in Arizona is ought to be taken into account when trying to detect or forecast the epidemic start in California. A further important remark is that the illustrated spatial spread is year-specific, and in other years rather different patterns may be observed.

\subsection{Contributions}
In this paper we formulate and analyze an epidemic detection problem within a multi-population paradigm. To do so, we develop a reduced compartmental model that extends the classical Susceptible-Infected-Recovered (SIR) setup to two population pools. Pools are interpreted as distinct geographic regions, e.g.~states or counties. To fix ideas, we consider the situation where the epidemic begins in Pool 1 and subsequently may be transmitted to Pool 2 via infecteds that travel between the two pools. The aim of the policy-maker is to detect, as soon as possible and in online fashion, the onset of epidemic in Pool 2.

To capture the inferential aspect, we assume that full information is available about the outbreak in Pool 1, but only partial information about Pool 2. As a result, one has to make imperfect decisions and in particular address the canonical trade-off between making announcements too early (so called ``false alarms'') and making decisions too late (``detection delay''). Indeed, if the detection is too late, then a certain number of infections would be missed and it would be harder to stop the epidemic from spreading. If the detection is premature, human, financial and reputational resources would be wasted. Therefore, a  careful trade-off between those costs should be done to balance costs due to epidemic morbidity and costs arising from policy actions. We then use the above cost analysis to quantify decision-making quality and to define optimality of detection strategies.

Mathematically, we cast the online detection problem as a dynamic optimization problem,  connecting to the classical dynamic programming formulation \cite{BertsekasBookVol1} in control theory. A major challenge with  dynamic programming (which is perhaps the prime reason for the lack in its uptake in the biosurveillance community) is computational bottlenecks due to the curse of dimensionality. Indeed,
the above optimization problem is nontrivial from several directions. First, because the underlying system is stochastic, the optimal solution is \emph{adaptive}, i.e.~a function of the current system state. Consequently, there is no simple description to the resulting detection strategy which is instead summarized through a detection \emph{map} that translates system states into optimal detection decisions. Second, the nonlinear dynamics of the SIR model preclude analytic solutions. Crucially, there are no analytic expressions for the future distribution of the system state, which necessitates the use of numerical approximations to solve the optimization problem. Third, because the system state is multivariate and too large to enumerate, the corresponding integrals are computationally demanding.

However, taking advantage of the detection strategy structure, which requires simply announcing at each stage whether the epidemic has reached Pool 2 or not, we implement an efficient numerical algorithm. Specifically, we rely on the recent Sequential Regression Monte Carlo (SRMC) method of \cite{GL14}, which blends modern statistical tools, including nonparametric regression and sequential design, with approximate dynamic programming, to drastically mitigate issues of computational efficiency.

The main contributions of this work are then threefold. First, we propose and analyze a multi-population extension of the classical SIR model, as well as a reduced version suitable for the Bayesian detection framework. Second, we develop and adapt an extension of the sequential regression Monte Carlo (SRMC) approach to efficiently solve the dynamic optimization problem. Third, we present a detailed investigation into the performance of the designed strategy, in particular in comparison to conventional threshold-based strategies.

The organization of the paper is as follows. Section \ref{sec:model} formalizes the mathematical aspects of our model, including the detection setup. The stochastic dynamics of the outbreak are rigorized in Section \ref{sec:sir}.
%
Section~\ref{sec:case-study} presents numerical illustrations of our method as well as comparison with other methods. Section~\ref{section SRMC} then describes the Sequential Regression Monte Carlo algorithm that we developed for our setup.
Section~\ref{section: discussion} provides the conclusion and the discussion on future extensions of our framework.

\subsection{Spatial Stochastic Epidemic Models}
Mathematical models of infectious disease epidemics have become an important tool in the arsenal of public health policy. In an idealized world, detection reduces to the mathematical problem of {clustering}, tracking the health status of the surveyed individuals and identifying unusual aberrations in either the temporal or spatial dimensions. In reality, there is the additional aspect of missing information which necessitates the application of statistical inference algorithms, as well as a mathematical model for the epidemic.
In the context of online inference, a simple mechanistic approach that allows for maximum tractability continues to be the most popular, and is also adopted here. Specifically, we rely on the formalism of an SIR model \cite{andersson} that implies proportional homogenous mixing between infecteds and susceptibles within a population pool. Spatial heterogeneity is captured by incorporating meta-populations, also known as patch models \cite{BallClancy93,Allen2009,Neal2012}. The multi-patch approach partitions the global population into distinct discrete regions or pools, allowing for local spread of the epidemic within each pool, as well as global transmission that is specified via a mobility matrix. As in \cite{BallClancy93,Neal2012} we assume that susceptibles are stationary, while infecteds can move or travel between the pools, creating cross-infections.


Alternative frameworks for epidemic spread include point process models \cite{Neill2011}, and network models \cite{Keeling2005} that provide more nuanced interaction between individuals to mimic existing social structures, such as households, schools, and workplaces. At even more detail, agent-based models \cite{BalcanEtal10} generate micro-simulations that provide a detailed synthetic view for each individual and their social interactions. Such models can also incorporate precise travel patterns \cite{Rvachev1985}. However, the latter paradigms are geared towards realistic \emph{forecasting} of epidemic progress and are less suited for online detection due to intractable inference in terms of observed data and the computational expenses in generating micro-scenarios.

A variety of approaches exist for constructing outbreak detection rules, see for example the recent survey by Shmueli and Burkom \cite{Burkom2010}, and the monograph by Lawson \cite{Lawson2013}. Quality control methods \cite{Cowl:Wong:Ho:Rile:Leun:meth:2006} introduced in the 1950s form the simplest class of rules and continue to be common. Other heuristics include moving-average tools \cite{Lawson2008}, various scan statistics \cite{spacetimeKulldorff,Neill2011}, and branching-process approximations \cite{Nishiura}. More explicit cost-benefit analysis for the trade-off between false alarms and detection delay can be applied using the Cumulative Sum (CUSUM) framework \cite{HadjiliadisLudkovski14}. CUSUM also underlies the early aberration response system (EARS) employed by the Centers for Disease Control \cite{Hutwagner2005}. Alternatively, Bayesian methods allow to further assess the uncertainty involved in decision-making based on partial information. Two main types are hidden Markov models \cite{LeStratCarrat,Mart:Cone:Lope:Lope:baye:2008} and Bayesian hierarchical models \cite{Chan2010,Sebastiani2006}. The Bayesian paradigm translates epidemic data into the posterior probability of an outbreak. To convert the latter into a detection rule, one typically employs a simple threshold strategy. For example, in \cite{Chan2010}, the authors recommend ``an alert for action if the posterior probability is larger than 70\%''. We further refine this approach by deriving \emph{optimal}, non-parametric detection strategies based on the inputted cost-benefit parameters.

Detection can be seen as a basic form of epidemic response, and indeed our computational methodology can be extended to this more general problem. In that sense, this paper extends the first author's previous work on stochastic control methods for controlling epidemics \cite{LudkovskiNiemi2011,LN10}. Similar to \cite{LudkovskiNiemi2011}, we design a Bayesian dynamic optimization algorithm for biosurveillance decision policy. Other mathematically oriented studies that consider optimal control of epidemics include \cite{TannerSattenspielNtaimo08,wrk}.

In the context of detection with limited information, a spatial epidemic model requires information fusion.
 Fusion of information channels for the purpose of biosurveillance has been an area of intense research in the past decade. On the one hand, novel information sources, such as social media \cite{Culotta2010} or internet data \cite{DukicPolsonPL} have created new opportunities for syndromic surveillance. On the other hand, developments in statistical fusion techniques \cite{Burkom2010,Bank:Datt:Karr:Lync:Niem:Vera:baye:2012,Noufaily2013} have led to new ways of integrating multivariate information streams. In particular, there has been a lot of interest in online Bayesian approaches \cite{lin2013sequential,Niemi2014,Nishiura,DukicPolsonPL,Shaman2014} that allow for predictive modeling and forecasting of epidemics. The above models all focus on a single homogenous population with the different surveillance channels complementing each other. In contrast, we consider multiple underlying population pools each with a distinct, but co-dependent information channel. In terms of explicitly accounting for spatial propagation, our work is closest to \cite{Ludkovski2012} who considered a spatial ``wave'' model for an epidemic. In the present article, we connect this framework to the SIR context, modeling epidemic spread across geographically-based population pools. The resulting decision strategy provides insights into integrating data from multiple spatial locales for the purposes of detection, cf.~Section \ref{section: discussion} below.

\section{Quickest Detection}\label{sec:model}

\subsection{Mathematical Model}
We work with a state-space model, denoting by $\mathfrak{X}_t$ the epidemic state at times $t=0,1,2,\ldots$. A typical length of one time period in biosurveillance is a week. The precise components of $\mX$ will be specified later; abstractly $\mathfrak{X}$ is taken to be a stochastic Markov process taking values in a state space $\cX \subset \mathbb{R}^d$, and summarizes information about both Pool 1 and Pool 2. In particular, $\mX$ contains information about the number of infecteds $I^{(k)}_t$ in Pool $k=1,2$ at time $t$. The transition kernel of $\mX$ is assumed to be time-stationary and is denoted by $p_s(\mx|\mathbf{y}) \equiv P( \mX_{t+s} = \mx | \mX_t = \mathbf{y})$, $\mx,\mathbf{y} \in \cX$.

The aim of the policy maker is to detect the onset of epidemic in Pool 2.
A detection strategy is probabilistically represented as a dynamic ``alarm'' which announces an outbreak in Pool 2, based on information gathered so far. Only a single announcement is allowed; once announced, the detection problem is assumed to be over.
 The set of such detection strategies is expressed through the set $\cS$ of $\cF$-stopping times, where $\mathcal{F}_t = \sigma (\mathfrak{X}_{0:t})$ is the information filtration generated by $\mathfrak{X}$ by time $t$. A strategy $\tau \in \cS$ is a random variable taking values in $\tau \in \{0,1,2,\ldots\}$, such that $\{ \tau = t \} \in \cF_t$ (this requirement captures the fact that $\tau$ must be ``online'' in terms of the information available so far). Thanks to the Markov property of $\mX$, the structure of $\tau$ can be summarized via a detection map. Indeed, at each time-step there is the binary decision to either
 ``announce'' an outbreak (subset $\mS$), or wait for another period (subset $\mathfrak{C}$). Since the evolution of $\mX$ is stationary in time, the corresponding partition of the state space is also independent of $t$. Dynamically, this implies that $\tau$ announces the epidemic the first time that the state $\mX$ enters the region $\mS \subset \cX$,
\begin{align} \label{eq: def.tau}
  \tau = \inf \{ t: \mX_t \in \mS \}.
\end{align}
Equation \eqref{eq: def.tau} gives a one-to-one correspondence between detection strategies $\tau$ and detection maps $\mS$. In other words, the detection strategies we consider are of online feedback type, based on the trajectory of $\mX$.

As mentioned, the dynamic optimization objective consists in optimally trading off the concern of premature announcements against any potential delays. These conflicting costs are measured through the immediate stopping cost $d(\mx)$ and the cost of waiting. The immediate costs are linked to the penalty for false alarms, specified by a given constant $C_{\text{FA}}$. We assume that $C_{\text{FA}}$ is paid if and only if the epidemic has not yet reached Pool 2, so that
\begin{align}
  \label{eq: cost_today}
d(\mx_0)&:= C_{\text{FA}} \cdot \mathbf{1}_{\{I_{0}^{(2)}=0\}}.
\end{align}
Waiting costs are assumed to be proportional to detection delay, i.e.~the time between the outbreak reaching Pool 2 and outbreak announcement. Define $\theta$ to be the time when the second population gets infected from the first population, i.e.
$$\theta := \inf \{t: I_{t-1}^{(2)}=0 \text{ and }I_t^{(2)}>0\}.$$
Then the detection delay is $\max(\tau - \theta,0)$ and carries cost $C_{\text{Delay}}\max(\tau - \theta,0)$. This is equivalent to charging waiting costs of $C_{\text{Delay}} \mathbf{1}_{\{I_t^{(2)}>0\}}$ at each step until surveillance is terminated at the random instant $\tau$, so that total waiting costs on $[0,\tau]$ are
\begin{align}
\label{eq: costs}
c(\mathfrak{X}_{0:\tau})&:=\sum_{s=0}^{\tau-1} C_{\text{Delay}} \mathbf{1}_{ \{I_s^{(2)}>0\}} + C_{\text{FA}} \mathbf{1}_{\{I_{\tau}^{(2)}=0\}}.
\end{align}
We will refer to the costs $d(\cdot)$ and $c(\cdot)$ as the immediate cost and the future cost, respectively.

\begin{remark}
Note that detection costs are intrinsically defined in terms of the count of infecteds in Pool 2, $I^{(2)}$, which is assumed to be unavailable to the policy-maker. Below we will operationalize \eqref{eq: cost_today} and \eqref{eq: costs} by taking conditional expectation with respect to information that is available, see \eqref{eq: costs_X}-\eqref{eq: cost_today_X}.
\end{remark}

The aim of outbreak detection is to pinpoint $\theta$, i.e.~ideally one takes $\tau = \theta$. However, this is not possible if only partial information is available about $\mX$, specifically about $I^{(2)}_t$. When $\tau$ and $\theta$ are different, $C_{\text{Delay}}$ penalizes the event $\{\tau > \theta\}$, and $C_{\text{FA}}$ penalizes $\{ \tau < \theta\}$. The cost structure in \eqref{eq: costs} is then a dynamic counterpart of the usual Type-I and Type-II errors in hypothesis testing.

 \subsection{Detection Problem}
\label{section detection}
Our detection problem is formalized as minimizing the expected future cost over all possible stopping times $\tau$ \cite{poor2009quickest}, i.e.~an optimal stopping problem. Namely, we define the value function $V$ as
\begin{equation}
V(\mx_0) :=\inf_{\tau \in S} E\left[ c (\mathfrak{X}_{0:\tau})|\mathfrak{X}_0=\mx_0 \right],
\label{eq: value_function}
\end{equation}
 where $\mx_0$ is the initial state.
 Assuming the infimum in~\eqref{eq: value_function} is achieved, the dynamic programming principle \cite{poor2009quickest} implies \begin{equation}
V( \mx_0)=\min \left( d (\mx_{0}), E\left[ V( \mathfrak{X}_{1})|\mathfrak{X}_0=\mx_0 \right] \right),
\label{eq: value_function2}
\end{equation}
where the conditional expectation operator is $$E[  V( \mathfrak{X}_{1})|\mathfrak{X}_0=\mx_0 ] = \int V(\mx) p_1(\mx | \mx_0) d\mx.$$
The minimum operator  in \eqref{eq: value_function2} corresponds to the idea that it is optimal to declare an outbreak if the immediate cost is smaller than the future cost, i.e.~the likelihood of false alarms is dominated by the cost of waiting. The former case is equivalent to the expectation of the value function at time 1 being greater than the immediate cost, and therefore we may classify the stopping region via
\begin{equation}
\mathfrak{S}:=\left\{\mx: E\left[ V( \mathfrak{X}_{1}) | \mathfrak{X}_0=\mx\right] -d(\mx) > 0\right\}.
\label{eq: stoppingregion0}
\end{equation}
Hence, in terms of  the above detection map, our goal is to optimally partition $\cX = \mS \cup \mathfrak{C}$ into two regions, such that $\mS$ consists of all initial states $\mx_0$ where it is optimal to declare the epidemic, and $\mathfrak{C}$ is its complement, where it is optimal to wait.

\subsection{Reduction to a Model Predictive Control Problem}
The characterization in \eqref{eq: value_function2} is implicit, since it features $V(\cdot)$ on both sides of the expression. Specifically, the value function $V$ corresponds to a fixed point \cite{BertsekasBookVol1} of the functional operator $\mathcal{L}$, defined by
$ (\mathcal{L} v) (\mx):= \min \left( d (\mx), E\left[ v( \mathfrak{X}_{1})|\mathfrak{X}_0=\mx \right] \right)$.
To solve for $V(\mx)$, a basic strategy is then to apply Picard-type fixed-point iterations. In other words, given some initial guess $V^{(0)}(\mx)$, we build a sequence of approximations via $ V^{(k)} := \mathcal{L} V^{(k-1)}$, or explicitly,
\begin{align}
\label{eq:mpc}
V^{(k)}( \mx_0)= \min \left( d (\mx_{0}), E\left[ V^{(k-1)}( \mathfrak{X}_{1})|\mathfrak{X}_0=\mx_0 \right] \right).\end{align}

 However, to guarantee the convergence of $V^{(k)}$ does not appear tractable, and the practical performance of \eqref{eq:mpc} is very sensitive to the initial guess $V^{(0)}$. To circumvent this challenge, we rely on the concept of model predictive control. To wit, we introduce an auxiliary parameter $t$ which can be intuitively thought of as \emph{forward time}. The value functions $V(t,\cdot)$ and detection maps $\mathfrak{S}_t$ are now also indexed by $t$.  We start with the trivial initial condition $V(0,\mx) := d(\mx)$, which corresponds to  $\mS_0 \equiv \cX$. Next, mimicking the classical dynamic programming on finite horizon, we define
%
%
\begin{equation}
V(t, \mx_0) :=\min \left( d (\mx_{0}), E\left[ V(t-1, \mathfrak{X}_{1})|\mathfrak{X}_0=\mx_0 \right] \right), \quad t=1,2,\ldots.
\label{eq: value_function3}
\end{equation}
Define the Q-value, also known as costs-to-go by
\begin{align}
\label{eq: Q}
q(t, \mx) := E[ V(t-1,\mX_1) | \mX_0 = \mx].
\end{align}
Then the stopping set at iteration $t$ is
\begin{equation}
\mathfrak{S}_t:=\left\{\mx_0 \in \cX : q(t,\mx_0) -d(\mx_{0})>0\right\}, \quad t=1,2,\ldots.
\label{eq: stoppingregion2}
\end{equation}
We may ``unroll'' the expectation encoded in $V(t-1,\mX_1)$ to write
\begin{equation}\label{eq:c-tau-star}
q(t,\mx_0) = E\left[ V(t-1, \mathfrak{X}_{1}) | \mathfrak{X}_0=\mx_0\right] = E\left[ c(\mathfrak{X}_{0:\tau^{(t)}}) | \mathfrak{X}_0=\mx_0 \right],
\end{equation}
where $\tau^{(t)} =  \min \{s \ge 1: \mathfrak{X}_{s} \in \mathfrak{S}_{t-s}\}$. This justifies the interpretation of $q(t,\cdot)$ as costs-to-go, since $c(\mathfrak{X}_{0:\tau^{(t)}})$ are indeed the future costs associated with \emph{not} stopping immediately.

\begin{figure}[!ht]
\centering
\includegraphics[width=\textwidth]{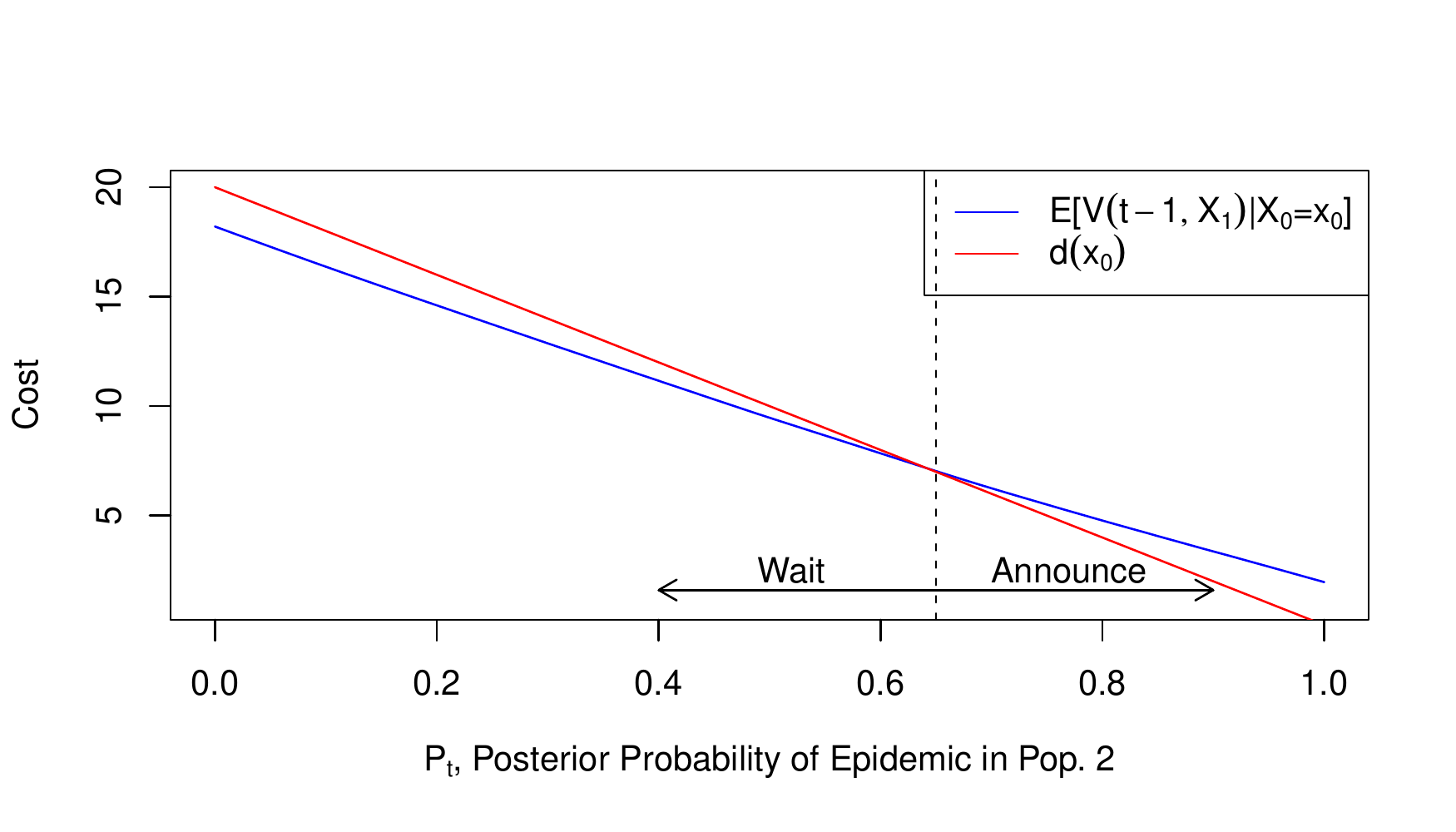}
\caption{Detection strategy at iteration $t=1$. The example is based on the model of Section \ref{sec:case-study}, with parameters in Table \ref{table: initial_values}. In the plot, the state of Pool 1 is held fixed at $S_0^{(1)}=1990, \ I_0^{(1)}=10$.}
\label{fig:scheme_costs}
\end{figure}

Figure~\ref{fig:scheme_costs} illustrates the first step of the recursion \eqref{eq: value_function3} at $t=1$. In the plot we compare
$$E\left[ V(0, \mathfrak{X}_{1}) | \mathfrak{X}_0=\mx\right] = E[ d(\mathfrak{X}_{1}) | \mX_0 = \mx]
$$
 against $d (\mx)$. As discussed, the epidemic is announced when $E\left[ V(t-1, \mathfrak{X}_{1}) | \mathfrak{X}_0=\mx\right] > d (\mx)$ (the right side of the plot).  In the opposite case (the left side of the plot), the optimal decision is to wait. As shown by the Figure, the structure of the decision map is driven by the regions where these two quantities are equal to each other, which corresponds to the detection \emph{boundary},
\begin{equation}
\partial \mathfrak{S}_t := \left\{\mx: E\left[ V(t-1, \mathfrak{X}_{1}) | \mathfrak{X}_0=\mx\right]= d (\mx)\right\}.
\end{equation}

The stopping region $\mathfrak{S}_t$ is our detection rule at iteration step $t$. It can be characterized as the optimal detection rule among all strategies in $\cS^{(t)} = \{ \tau \in \cF : \tau \le t \}$ that are upper-bounded by $t$ (By construction, $\tau^{(t)} \le t$). As $t \to \infty$, we have that the set of admissible rules expands $\cS^{(t)} \nearrow \cS$, and hence we expect that $\mathfrak{S}_t \to \mS$ and $V(t,\mx) \to V(\mx)$ . Intuitively, for large $t$, the recursively defined \eqref{eq: value_function3} converges to a stationary case that ought to be the fixed point defining $V(x)$ in \eqref{eq: value_function}. The above convergence can be improved via model predictive control (also known as receding horizon control) \cite{Nevistic96modelpredictive} which applies the fixed detection map $\hat{\mathfrak{S}}^{(k)}$, rather than the time-dependent $\hat{\mathfrak{S}}_t$ at each step, cf.~Section \ref{sec: RMC}.



\section{Epidemic Model}\label{sec:sir}
\subsection{Multiple Population SIR model}
\label{def: SIR}
A susceptible-infected-recovered (SIR) model provides an aggregate ``gravity'' view of the epidemic by focusing on three basic types of individuals in the population: susceptible, infected and recovered. Susceptible individuals are the ones who haven't experienced the disease yet. Interaction between an infected and susceptible individuals can lead to an infection. Thus, contacts stochastically generate new infecteds who in turn can further infect other susceptible individuals. After some time an infected individual recovers and becomes immune (i.e.~becomes a Recovered): he/she can no longer infect others or get infected.

While the detection problem is specified at the discrete instances $t=1,2,\ldots$, for describing outbreak dynamics it is more convenient to work with continuous-time dynamical systems. As in \cite[Ch.~6]{andersson}, we thus first recall the multi-type stochastic SIR model in continuous time. The overall epidemic state at epoch $t \in \mathbb{R}_+$ is denoted by $\{\mathbf{S}_t, \mathbf{I}_t,  \mathbf{R}_t\}$, where $\mathbf{S}_t=\{S^{(1)}_t, \ldots, S^{(K)}_t\}$, $\mathbf{I}_t=\{I^{(1)}_t, \ldots, I^{(K)}_t\}$ and $\mathbf{R}_t=\{R^{(1)}_t, \ldots, R^{(K)}_t\}$ are vectors denoting the count of susceptible, infected  and recovered individuals in each of $1\leq k\leq K$ meta-populations. We assume that the pool size of each meta-population is fixed at  $M^{(k)}=S_t^{(k)}+ I_t^{(k)}+R_t^{(k)}$.
As a result, we omit further mention of $R_t^{(k)}$ since  it can be found from $R_t^{(k)}=M^{(k)}-S_t^{(k)}- I_t^{(k)}$.

The continuous evolution of the state process
 $\{\mathbf{S}_t, \mathbf{I}_t\} \in \{ (\mathbf{s},\mathbf{i}) : s_k+i_k \le M^{(k)} \;\forall k \le K\}$ is described through Markov chain or stochastic kinetic system language. Namely, the epidemic state is piecewise constant in time. Next, there are $2K+K(K-1)$ possible transitions, described by the reaction channels \cite{WilkinsonBook}:
\begin{equation}
\left\{
\begin{array}{r r l l}
\text{Infection} & S^{(k)}+ I^{(k)} &\to 2I^{(k)}  & \text{w/rate} \quad \beta_k I^{(k)} \frac{S^{(k)}}{M^{(k)}}\\
\text{Transmission} &  S^{(k)}+I^{(k')} &\to I^{(k)} + I^{(k')} &\text{w/rate} \quad \beta_{k,k'} I^{(k')} \frac{S^{(k)}}{M^{(k)}} \\
\text{Recovery} &  I^{(k)} &\to\emptyset &\text{w/rate} \quad \gamma I^{(k)}
\end{array}
\right\}
\label{eqn: reaction_channels}
\end{equation}
where each reaction is further indexed by $1\leq k\leq K$ and $k'\leq K, \ k'\neq k$.

 \begin{figure}[t]\centering   
\begin{tikzpicture}[->,>=stealth',shorten >=1pt,auto,node distance=2cm,
                    thick,main node/.style={circle,draw,font=\sffamily\bfseries}]

  \node[main node] (1) {$S_t^{(1)}$};
  \node[main node] (2) [below of=1] {$I_t^{(1)}$};
  \node[main node] (3) [below of=2] {$R_t^{(1)}$};
  \node[main node] (4) [right of=1, xshift=40mm] {$S_t^{(2)}$};
  \node[main node] (5) [below of=4] {$I_t^{(2)}$};
  \node[main node] (6) [below of=5] {$R_t^{(2)}$};

  \path [every node/.style={font=\sffamily\small}]
    (1)  edge  node[ left] {$\beta_1 I_t^{(1)}\frac{ S_t^{(1)}}{M^{(1)}}$} (2)
    	 edge  [bend left] node[right, pos=0.4] {$ \beta_{1,2} I_t^{(2)}  \frac{S_t^{(1)}}{M^{(1)}}$} (2)
    (2)  edge  node[left] {$\gamma I_t^{(1)}$} (3)
          edge [bend left=15] node [right] {}(5.120) %
    (4)  edge  node[right] {$\beta_2 I_t^{(2)}\frac{ S_t^{(2)}}{M^{(2)}} $} (5)
          edge [bend right] node[ left, pos=0.4] {$ \beta_{2,1} I_t^{(1)}  \frac{S_t^{(2)}}{M^{(2)}}$} (5)
    (5)  edge  node[right] {$\gamma I_t^{(2)}$} (6)
           edge [bend right=15] node [left] {}(2.60);

   \tikzset{blue dotted/.style={draw=blue!50!white, line width=1pt,
                               dash pattern=on 1pt off 4pt on 6pt off 4pt,
                                inner sep=4mm, rectangle}};
    \draw[-,blue dotted]  (-2,1) -- (2.5,1) -- (2.5, -5) -- (-2,-5) -- (-2,1) ;
    \draw[-,blue dotted] (3.5,1) -- (8,1) -- (8, -5) -- (3.5,-5) -- (3.5,1) ;
\end{tikzpicture}
\caption{Stochastic SIR epidemic model in two populations.}
\label{fig:SIR2}
\end{figure}
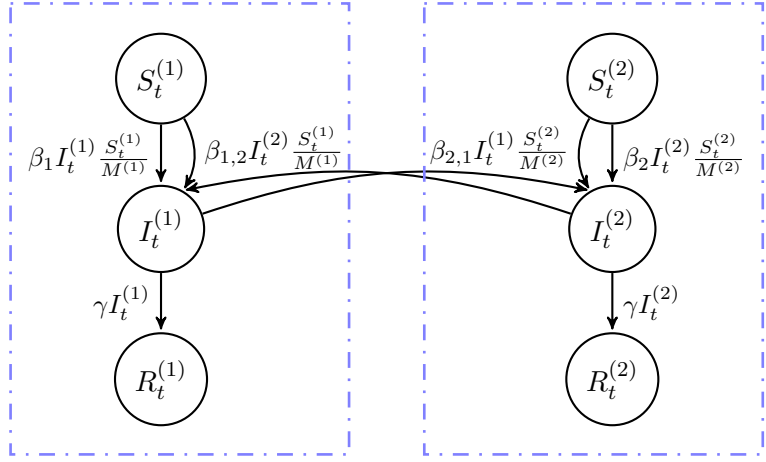

The first transition represents an infection of a susceptible individual by an infected individual from the same pool $k$.  This transition happens at rate $\beta_k I_t^{(k)} \frac{S_t^{(k)}}{M^{(k)}},$
where $1\leq k \leq K$ and $\beta_k$ is a contact rate of  infected and susceptible individuals within the $k$-th meta-population.

The second transition is a transmission: an infection of a susceptible individual from pool $k$ by an infected individual from a different pool $k'$. The frequency of such infections is
$\beta_{k,k'} I_t^{(k')} \frac{S_t^{(k)}}{M^{(k)}},$
where $1\leq k, k' \leq K$ and $\beta_{k,k'}$ is a contact rate of  infected and susceptible individuals from different populations.
Since contacts between individuals from different populations are less frequent, $\beta_{k,k'} \ll \beta_{k'}$. To reduce the number of parameters, we thus assume that cross-population interactions occur at rate $\beta_{k,k'} \equiv \alpha \beta_{k'}$, where $\alpha$ is the proportion of ``travelers'' in each pool. Thus, cross-contacts happen at the fraction $\alpha$ of a contact rate within one population; a typical range is $\alpha \in [0.01,0.2]$.

The last transition in \eqref{eqn: reaction_channels} is a recovery and subsequent immunity of an infected individual in a population $k$. The rate of transition is $\gamma I_t^{(k)},$ where $\gamma$ is a recovery rate, independent of the pool index $k$. This can be interpreted as individuals staying infected for an Exponentially distributed time with mean $1/\gamma$.

In this paper we focus on two-population models, positing that the outbreak begins in Pool 1 and may subsequently spread to Pool 2, where it is to be detected. Accordingly, we will be fusing information from Pool 1 and Pool 2 to identify the onset of epidemic in Pool 2. We assume that the two pools have similar characteristics, so that all parameters are homogenous in $k=1,2$. Thus, the two-population SIR model, shown in Figure \ref{fig:SIR2} has 3 parameters  $\alpha$ -- the mixing parameter between two populations, $\beta$ -- the within-pool contact rate of infected and susceptible individuals, and $\gamma$ -- the recovery rate. These parameters are assumed to be known and are a function of the modeled disease family (e.g.~influenza or dengue fever), the demographics and public health characteristics of the populations, and the travel patterns across pools.

\subsection{Partial Observation}


Under full observations the detection problem \eqref{eq: value_function} would be trivial, since one can directly track $I^{(2)}_t$ and declare an outbreak as soon as there any infecteds in the second pool. However, realistically $I^{(2)}$ is not observed. Some of the reasons include mis-diagnoses among infecteds, patients not seeking care, false positives, mis-reporting or lack of reporting of epidemiological data, etc. Consequently, we assume that the true size of the S/I/R compartments in Pool 2 is not known. To simplify the presentation, we assume that $I^{(1)}_t$ \emph{is observed} in Pool 1, perhaps due to better epidemiological surveillance in that pool.

In our detection problem, the main event of interest is the presence of any infecteds in Pool 2, $\{ I^{(2)}_t > 0 \}$. Accordingly, we consider
 $\tilde{P}_t = P( I^{(2)}_t > 0 | \cG_{t})$, the posterior probability that the epidemic started in the second population given the limited knowledge about it available by time $t$, here summarized by some information set $\cG_t$. Depending on assumptions about the observations structure, $\tilde{P}_t$ may be available in closed form (e.g.~through Bayesian conjugate updating \cite{LN10,lin2013sequential}) or may have to be only approximately computed through e.g.~particle filtering methods \cite{Niemi2014,Skvortsov2012}. The latter method, which computes the whole posterior distribution $\pi_t \sim I^{(2)}_t | \mathcal{G}_t$, is computationally expensive, while conjugate updating requires carrying several sufficient statistics about the posterior of $I^{(2)}_t$. In either case, $\tilde{P}_t$ on its own is not Markovian, and hence does not possess simple dynamics. Therefore we propose a model that works with a simplified, Markovian version of $\tilde{P}_t$, which we denote as $P_t$.


\begin{figure}[ht]
\begin{center}
\includegraphics[width=0.9\textwidth,trim=0.2in 0.2in 0.2in 0.2in]{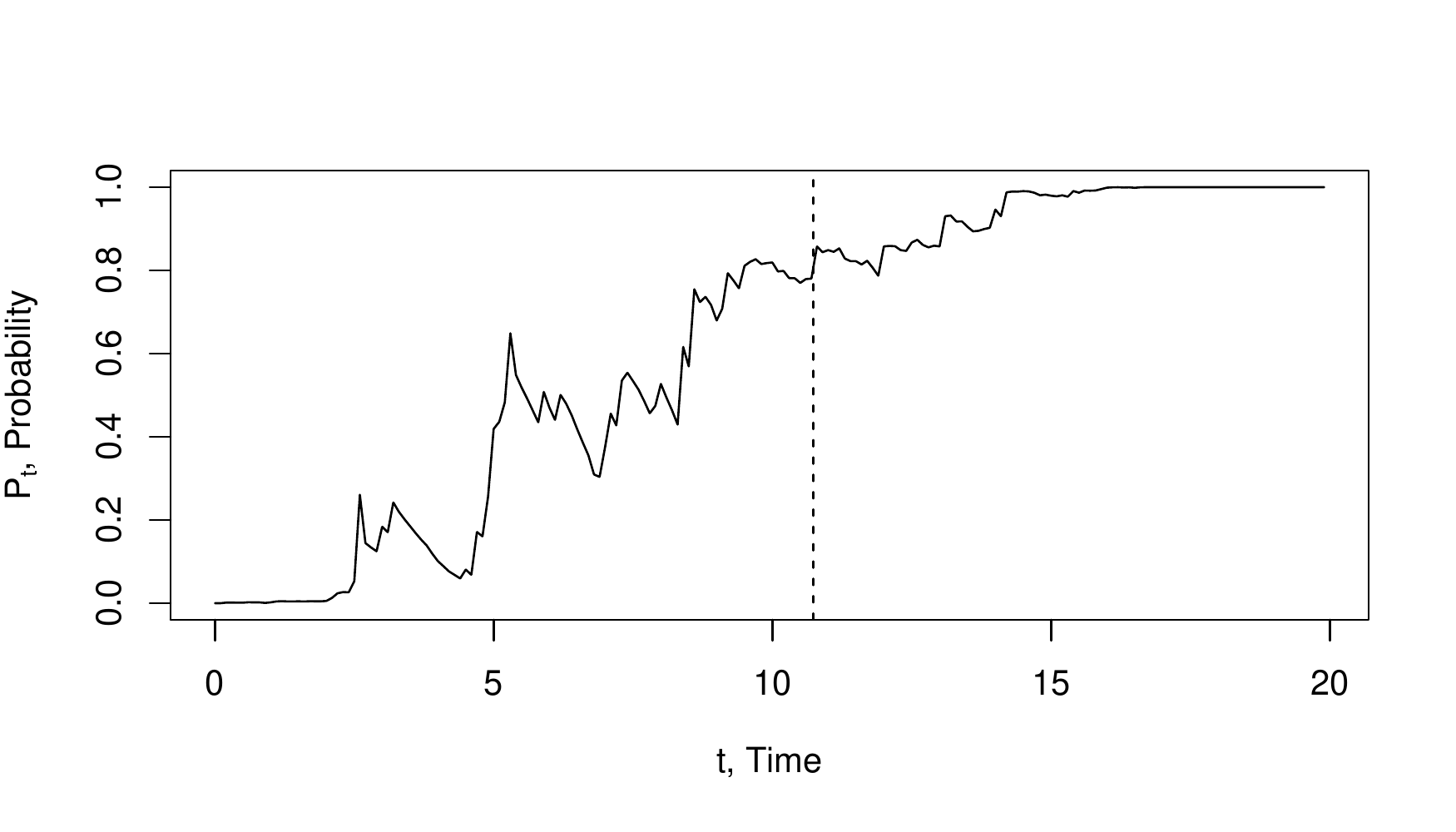}
\end{center}
\caption{Posterior probability that the epidemic started in the second population $\tilde{P}_t$ vs.~time $t$, where the dashed line represents the actual start time $\theta$ of outbreak in Pool 2. The plot was constructed using particle filtering using the following model parameter values: $\beta=0.75, \ \alpha=0.01, \ \gamma=0.5, \ M^{(1)}=M^{(2)}=2000$.}
\label{fig: post_prob}
\end{figure}

Figure~\ref{fig: post_prob} shows a sample scenario of the evolution of $\tilde{P}_t$ in a partially observed framework. The plot was generated using particle filtering and used the two-pool model \eqref{eqn: reaction_channels} with noisy Poisson-type observations in each pool \cite{Shatskikh16}. We observe that $\tilde{P}_t$ tends to drift up (i.e.~posterior probability of outbreak increases over time) and eventually hits 1.


\subsection{Reduced Model}
\label{section reduced model}

Our reduced model consists of  the state of epidemic in the first population  $\left\{S_t^{(1)}, I_t^{(1)}\right\}$ and a process $P_t$ that is interpreted as the probability that the epidemic reached Pool 2 conditional on the information $\cG_t = \sigma( S_{0:t}^{(1)}, I_{0:t}^{(1)})$ from Pool 1.  The first two components $S_t^{(1)}$ and $I_t^{(1)}$ come from a one-population SIR model (see the definition of SIR model for $K=1$ in Section~\ref{def: SIR}). To prescribe the dynamics of the pseudo-posterior $P_t$,
%
%
we decompose the event $\{ I^{(2)}_t > 0 \} \equiv \{ \theta \le t \}$ into two cases: the event that the epidemic already started at time $t-1$ (i.e.~$\theta \leq t-1$), and the event that it starts at $t=\theta$. We also add some stochastic noise to denote exogenous fluctuations in our posterior estimates regarding the second pool. In total, we thus assume that
\begin{align}\label{eq: second_term}
P_t = P_{t-1} + P(I_{t-1}^{(2)}=0 \text{ and }I_t^{(2)}>0|\cG_{t-1}) + \delta_t,
\end{align}
where $\delta_t$ are i.i.d.~noise terms. Intuitively, the probability of outbreak has a positive drift over time, and the drift is precisely the posterior probability of the outbreak beginning during the current period, $\{\theta \in [t-1,t]\}$.

 From the SIR dynamics, the probability that $\{\theta \in [t-1,t]\}$ conditional on Pool-1 observations up to previous stage  $t-1$, is equal to the probability that an infected from Pool 1 interacts with a susceptible from Pool 2, times the conditional probability that $\{ \theta > t-1\}$. The former happens with rate $\alpha \beta I_{s}^{(1)}\frac{S_{s}^{(2)}}{M^{(2)}}$, $s \in [t-1,t]$, while the latter event is the complement of $\{ \theta \le t-1\}$ and hence has probability $1-P_{t-1}$. Using the fact that conditional on $\{\theta \ge t\}$, $M^{(2)}=S_{t-1}^{(2)}$, and making the transition rate constant on $[t-1,t]$ we obtain
\begin{equation}
P(\theta \in [t-1,t]|\cG_{t-1}) \simeq  \alpha \beta I_{t-1}^{(1)} (1-P_{t-1}).  \label{eq: prob_new_infect_simplified}
\end{equation}

To guarantee $P_{t} \in [0,1]$ is interpretable as probability we confine it to $[0,1]$, yielding
\begin{align}\label{eq:reduced}
{P_{t}:= \begin{cases}
 0\vee (P_{t-1} + \alpha \beta I_{t-1}^{(1)}(1-P_{t-1}) + \delta_t ) \wedge 1, &\text{if } P_{t-1} \neq 1\\
 1, &\text{if } P_{t-1} = 1
 \end{cases}}\end{align}
In our simulations we use centered Gaussian noise $\delta_t\overset{\text{i.i.d}}{\sim} \mathcal{N}(0, \sigma^2_{\delta})$ with variance $\sigma^2_{\delta}$, however it can take any distribution. Note that $P_{t}=1$ is an absorbing state, representing certainty that the outbreak reached Pool 2, while $P_t = 0$ is a boundary case, since even if it is certain that the outbreak is currently not in Pool 2, it can still get cross-infected in the future. Similar features hold for the true posterior probability $\tilde{P}_t$, cf.~Figure~\ref{fig: post_prob}.
 Alternative models for probability of outbreak $P_t$, are discussed in Section~\ref{section: discussion}.

 \begin{remark}
 Note that \eqref{eq:reduced} is in discrete-time; to connect to the continuous-time dynamics of SIR one could take the limit as the time increment goes to zero, obtaining a diffusive model $dP_t = \alpha \beta I^{(1)}_t (1-P_t) \,dt + \delta dW_t$ where $(W_t)$ is a Brownian motion. However, since detection is assumed to take place only at instances $t=1,2,\ldots$, we prefer to work with \eqref{eq:reduced} as is.
\end{remark}




\subsection{Detection within the Reduced Model}

\begin{figure}[ht]
\begin{center}
\includegraphics[width=0.9\textwidth,trim=0.2in 0.2in 0.2in 0.2in]{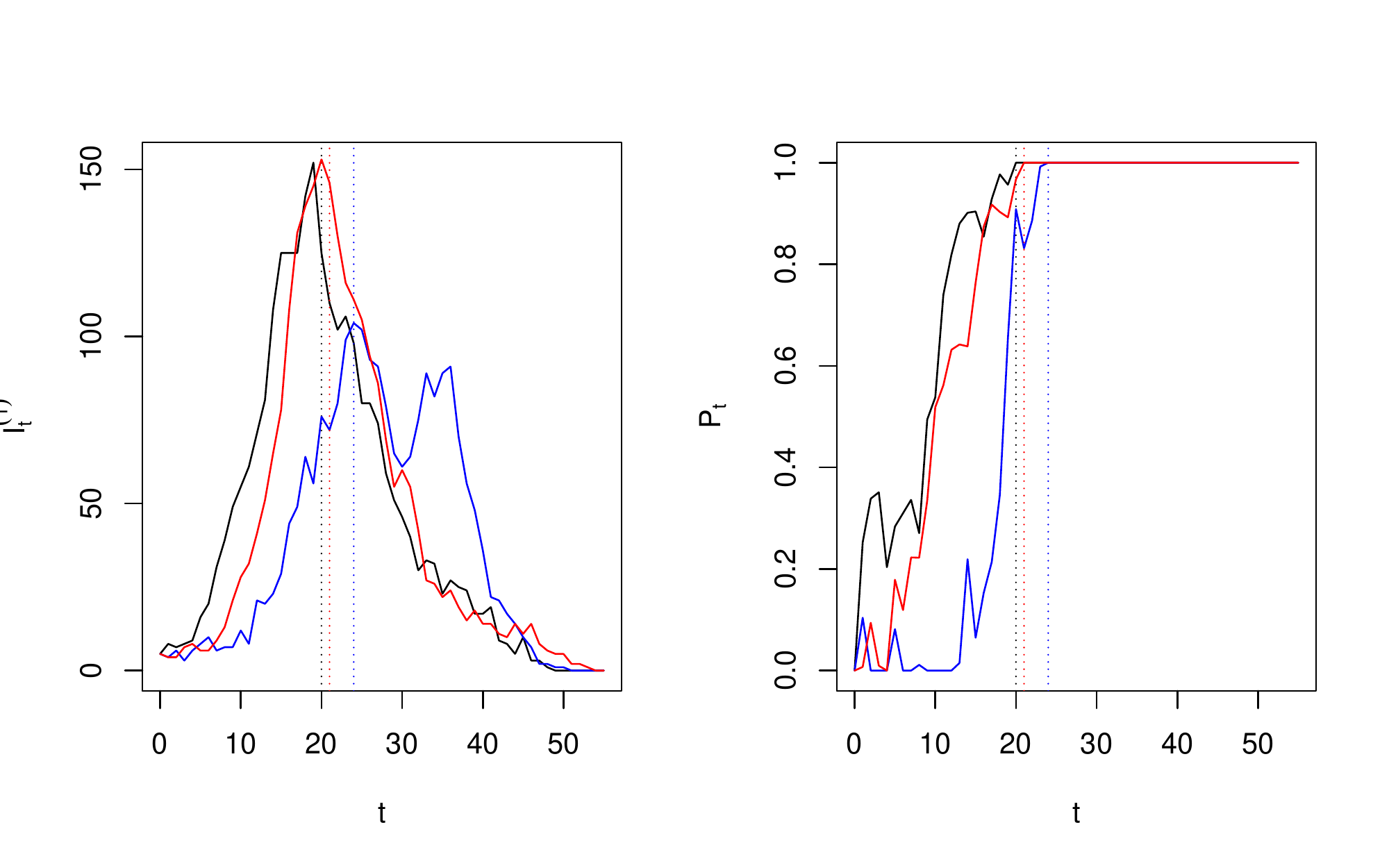}
\end{center}
\caption{Three sample trajectories of $\mathfrak{X}$ with the initial condition $S_0^{(1)}=1995$, $I_0^{(1)}=5$, $P_0=0$ and  outbreak parameters from Table \ref{table: initial_values}. Left panel is the plot of $\{I_t^{(1)}\}$, the number of infecteds in the first population, and right panel is the plot of $\{P_t\}$, the posterior probability that the epidemic started in the second population. The vertical dotted lines represent times when $P_t$ hits 1 and outbreak becomes certain.}
\label{fig: process_IP}
\end{figure}

To sum up, the developed reduced 2-pool model has a 3-dimensional state $\{\mathfrak{X}\}_t=\left(S_t^{(1)}, I_t^{(1)}, P_t\right)$ with state space
$$
\cX := \{ (s,i,p) : s,i \in \mathbb{N}, s+i < M^{(1)}, p \in [0,1] \}.
$$
Figure \ref{fig: process_IP} shows a few sample trajectories of $\mX$ to illustrate the resulting dynamics.

Our detection problem \eqref{eq: stoppingregion2} relies on the computation of the immediate and future expected costs $E\left[c(\mathfrak{X}_{0:\tau}) | \mathfrak{X}_{0}\right]$ and $d(\mathfrak{X}_0)$. Re-writing the definitions of  immediate and future costs \eqref{eq: cost_today} and \eqref{eq: costs}  in terms of the event $\{ I^{(2)}_t > 0 \}$, and taking conditional expectation we obtain:
\begin{align}
\label{eq: cost_today_X}
d(\mathfrak{X}_0) &:= C_{\text{FA}}(1-P_0),\\
\label{eq: costs_X}
c(\mathfrak{X}_{0:\tau})  &:=\sum_{s=0}^{\tau-1} C_{\text{Delay}}P_s + C_{\text{FA}} (1-P_{\tau}),
\end{align}
where $\tau \in \mathcal{S}$. Rather than in terms of the unobserved $I^{(2)}$, the above expressions are now given in terms of the component $P_t$, allowing to measure detection costs within the $\mX$-model. Notice that $d(\mX_0)$  is a function of $P_0$ and  $c(\mathfrak{X}_{0:\tau})$ is a function of the future trajectory $\{P_s, s=0,\ldots, \tau\}$.

Our goal is to find the detection maps  $\hat{\mathfrak{S}}_t$ for $t=1,2,\ldots,$ defined recursively in \eqref{eq: stoppingregion2}. To do so, at each step we need to evaluate $E\left[ V(t-1, \mathfrak{X}_{1}) | \mathfrak{X}_0=\mx\right]$ and $d(\mx)$. The immediate cost $d(\mx)$ can be computed exactly via~\eqref{eq: cost_today_X}. However, the expectation $E\left[ V(t-1, \mathfrak{X}_{1}) | \mathfrak{X}_0=\mx\right]$ can not be computed analytically since there are no closed-form expressions for the distribution of $\mathfrak{X}_{0:\tau}$. In Section \ref{sec: SRMC} we present the sequential Regression Monte Carlo approach which offers an efficient way to empirically estimate $\hat{\mathfrak{S}}_t$ based on synthetically generated epidemic scenarios. We then use Model Predictive Control to estimate the stationary detection map ${\mathfrak{S}}$.


\section{Case Study}
\label{sec:case-study}
To illustrate the dynamic detection strategy within our 2-pool model, in this section we present a detailed case study. Table~\ref{table: initial_values} summarizes the parameters used. Epidemic parameters are taken to be $\beta = 0.75$ and $\gamma=0.5$. Thus, the initial reproduction ratio is $\mathcal{R}_0 = \beta/\gamma = 1.5$, which is a moderately infectious epidemic.  We assume that the pool mixing parameter is $\alpha=0.01$, which is reasonable for pools representing well-separated cities or counties. The inference noise  in \eqref{eq:reduced} is taken to be Gaussian with variance $\delta_t \sim \mathcal{N}(0, \sigma^2_\delta=0.01^2)$.
For the detection costs in \eqref{eq: cost_today_X}-\eqref{eq: costs_X}, we take without loss of generality $C_{\text{Delay}} = 1$ and fix $C_{\text{FA}} = 20$. As we will see, this corresponds to a moderate penalty for false alarms.

\begin{table}[ht]
\centering
\begin{tabular}{c | l l l  }
\hline
Epidemic: 			&  $M^{(1)}= 2000$			& $S_{0}^{(1)}= M^{(1)} - I_{0}^{(1)}$ &$\sigma_{\delta}=1/100$\\
				&$\beta = 0.75$		& $\alpha = 0.01$			& $\gamma = 0.5$ \\
Costs/Penalties:  & $C_{\text{FA}} = 20$ 	&$C_{\text{Delay}} = 1$ 		&  \\
\hline
\end{tabular}
\caption{Outbreak and costs parameters for the case study of Section \ref{sec:case-study}. $\sigma_\delta$ refers to the noise in $P$, cf.~\eqref{eq:reduced}.}
\label{table: initial_values}
\end{table}

 \begin{figure}[ht]
\centering
\includegraphics[width = 0.85\textwidth]{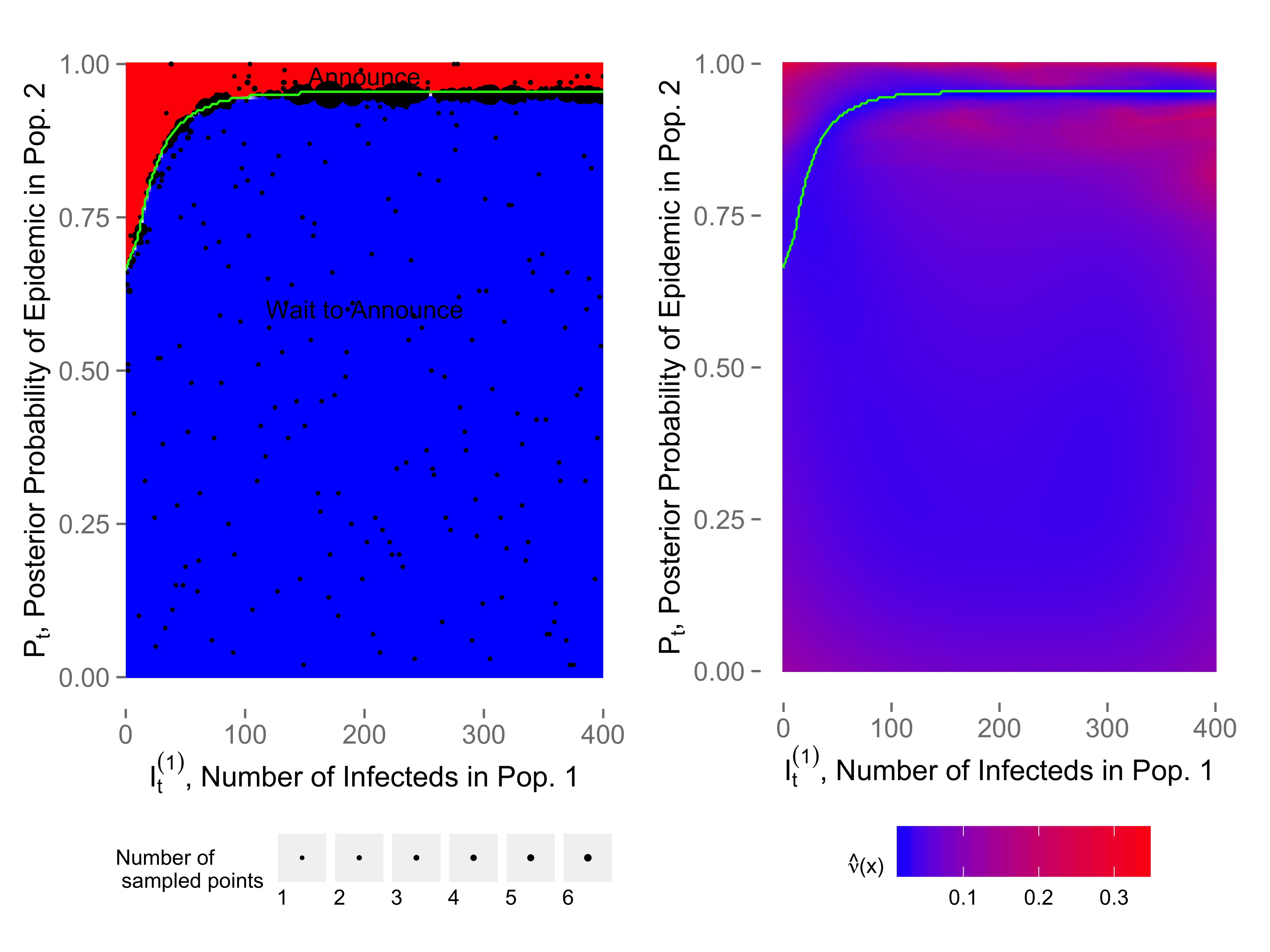}
\caption{Left panel: detection rule $\mS^{LP}_{20}$ in terms of $I^{(1)}$ and $P$. The detection boundary $\partial \mathfrak{S}^{LP}_{20}$ is shown with the solid curve. We also show the experimental design $\cZ$ that was used, illustrated with the scatterplot. Size of pixels corresponds to the number of times that neighborhood was sampled. Right panel: standard errors $\hat{v}(\mx)$ from \eqref{eq: variance_fitted}.
Observe lower standard errors in regions where the design $\cZ$ is more dense.
\label{fig: sample_detection_map}}
\end{figure}


 So far the case study features a three-dimensional state $\{ S^{(1)},I^{(1)},  P \}$, so that the resulting detection maps are in 3-D. To aid visualization, we consider a variant with a reduced dimension. Namely, we drop the component $S^{(1)}$ measuring the number of infecteds in Pool 1. Indeed, at the early stages of the outbreak the ratio $S_t^{(1)}/M^{(1)}$ is approximately one. As a result, one may assume that the rate of infections in Pool 1 is simply $\beta I^{(1)}_t$, which corresponds to the classical branching process epidemic model \citep{andersson}. It is known \cite{ball19951} that this approximation remains valid up to  $t = O( \log( M^{(1)})$ by which time, $I^{(1)}_t = O( \sqrt{M^{(1)}})$.; therefore it works especially well in large populations, and hence is termed a large-population (LP) approximation. The LP model only has two dimensions, $\mX' := \{ I^{(1)}, P \}$ allowing to plot the corresponding 2-D stopping set $\mS^{LP}$.

Figure~\ref{fig: sample_detection_map} shows $\mS^{LP}$ generated under the conditions of Table~\ref{table: initial_values} and the above large population assumption.
As expected, epidemic detection is triggered once the posterior probability $P_t$ of $\{I^{(2)}_t>0\}$, is high enough. However, we observe that detection is also highly sensitive to values of $I^{(1)}_t$; for instance detection is progressively delayed as $I^{(1)}_t$ gets bigger. This dependence between the two pools in terms of decision making illustrates the underlying cross-pool information fusion. Intuitively, detection should take place once $P_t$ is high enough. However, conditional on a fixed $P_t$,  larger number of Pool 1 infecteds makes an impending outbreak in Pool 2 more likely, lowering waiting costs. Hence, the detection boundary curves in $I^{(1)}$. Mathematically, recall that in \eqref{eq:reduced}, the growth rate of $P$ increases in $I^{(1)}$. As a result, for large values of $I^{(1)}_t$, one may expect that the next-stage $P_{t+1}$ will also be large, i.e.~move into  the ``Announce'' region quicker. This again lowers the waiting costs and therefore delays announcement.

\subsection{Evaluating Detection Rules}

\begin{figure}[ht] \centering
\includegraphics[width=0.65\textwidth,height=2in,trim=0.2in 0.2in 0.2in 0.2in]{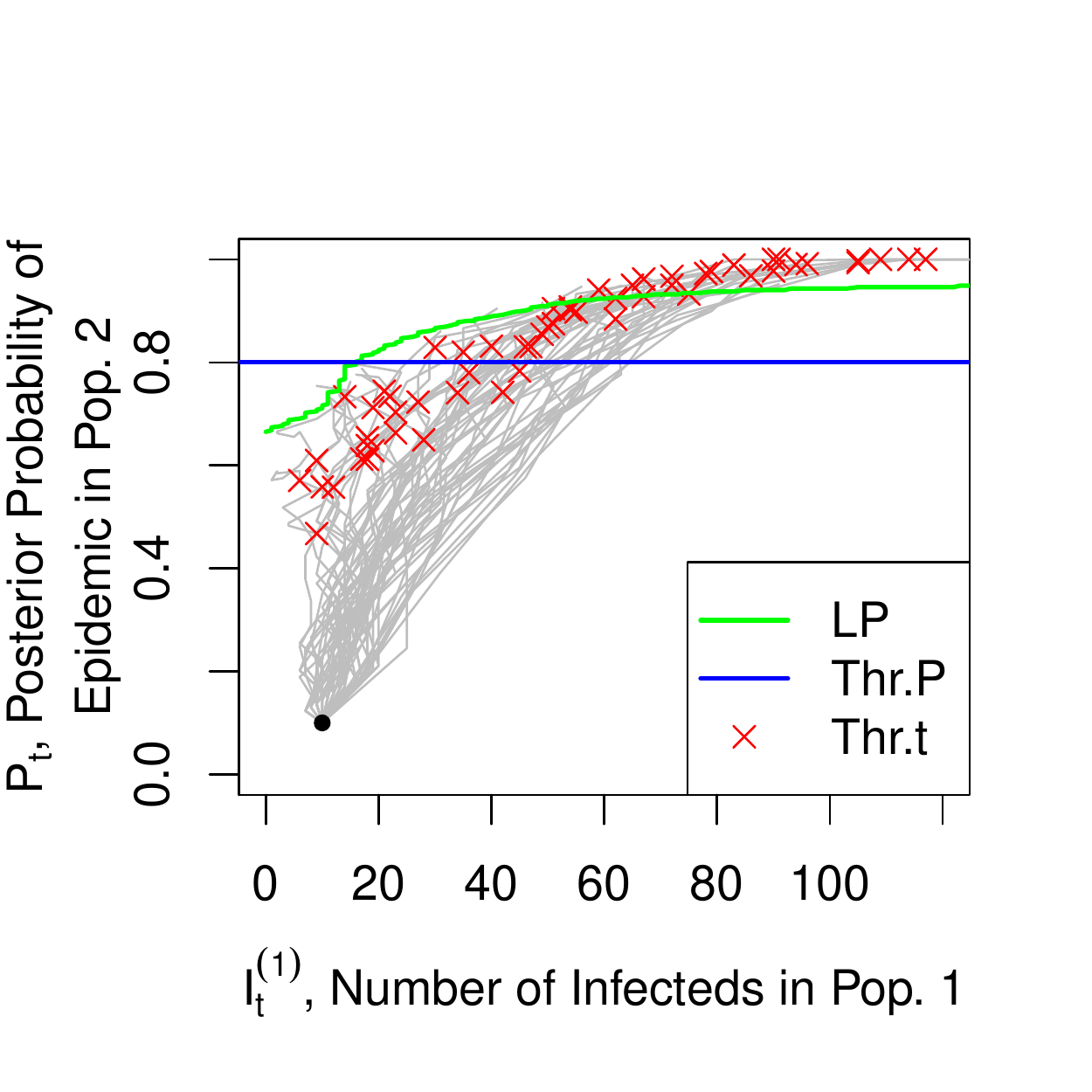}
\caption{
 Fifty sampled epidemic trajectories $\{I^{(1)}_t, P_t\}, t=1,\ldots, \tau$ emanating from the initial state $I_0^{(1)}=10$ and $P_0=0.1$. We show the LP detection boundary (namely $\partial \mathfrak{S}^{LP}_{20}$), as well as a threshold strategy that announces epidemic as soon as $P_t \ge \bar{P} = 0.8$. Lastly, the red crosses denote the locations of the trajectories at $t=8$, which is the basis of the alternate Threshold-t strategy. \label{fig: trajectories}}
\end{figure}

Figure~\ref{fig: trajectories} shows dynamic decision-making in the LP model through a collection of generated trajectories of $\mX' = \{I^{(1)}_t, P_t\}$ and their corresponding detection times $\tau^{LP}$, the first time the state process $\mX'$ enters the stopping set $\mS^{LP}$. We observe that the trajectories  generally move north-east, as both $P$ and $I^{(1)}$ tend to increase. However, the rate at which they grow and the precise direction are uncertain and vary across scenarios. Consequently, at detection, both $P_{\tau^{LP}}$ and $I^{(1)}_{\tau^{LP}}$ have a nontrivial distribution.

To better understand the detection map $\mS^{LP}$, we analyze the resulting detection strategy given by $\tau^{LP}$ and compare it to alternatives. Two classes of simpler  detection rules are Threshold-P and Threshold-t. The Threshold-P strategy announces an outbreak as soon as $P_t \ge \bar{P}$ for a given threshold $\bar{P}$. Hence, it acts solely based on local (posterior) information about Pool 2. This mimics the CDC policy \cite{Hutwagner2005} of announcing an epidemic when the number of infecteds in Population 2 crosses some pre-specified level. In contrast to the fused detection strategy with a curved detection boundary which jointly takes into account both $P_t$ and $I^{(1)}_t$, Threshold-P rule only uses $P_t$ for detection decisions, yielding a flat, horizontal detection boundary in Figure \ref{fig: trajectories}. The threshold-t strategy is a simple non-adaptive strategy that announces at the fixed stage $\bar{t}$. It is illustrated in Figure \ref{fig: trajectories} where we record the joint distribution of $I^{(1)}_{\bar{t}}, P_{\bar{t}}$ at $\bar{t} = 8$.


\begin{table}[hb]
\begin{tabular}{r|cc|cc|r}
				&\multicolumn{2}{|c|}{Detection time $\tau$}	& \multicolumn{2}{|c|}{Realized Cost $Q$} 	& \multirow{2}{*}{PFA $E[1-P_{\tau}]$}   \\ 
					& Mean 	& StDev.		& Mean 	& StDev.			&   \\ \hline\hline
Optimal  				& 8.86  	& 2.59		& 6.53	& 1.70 			& 8.2\% \\
LP					& 9.32   	& 2.95		& 6.57	& 1.81  			& 6.4\% \\
Threshold-P   			& 7.88   	& 2.85		& 7.03	& 1.58  			& 15.3\% \\
Threshold-t    	   		& 8.00     	& N/A   		& 7.18 	& 2.21 			& 14.4\%  \\  \hline
\end{tabular}
\caption{Comparison of Optimal, Large Population(LP), Threshold-P with $\bar{P}=0.8$ and Threshold-t with $\bar{t}=8$ strategies. Statistics are based on 1000 synthetic trajectories of $\{I^{(1)}, S^{(1)}, P\}$, where $Q = c(\mathfrak{X}_{0:\tau^{(t)}})$.}
\label{table: comparison_strategies}
\end{table}

\begin{figure}[ht]
\centering
\begin{tabular}{cc}
\includegraphics[width=0.47\textwidth,trim=0.25in 0.25in 0.4in 0.2in]{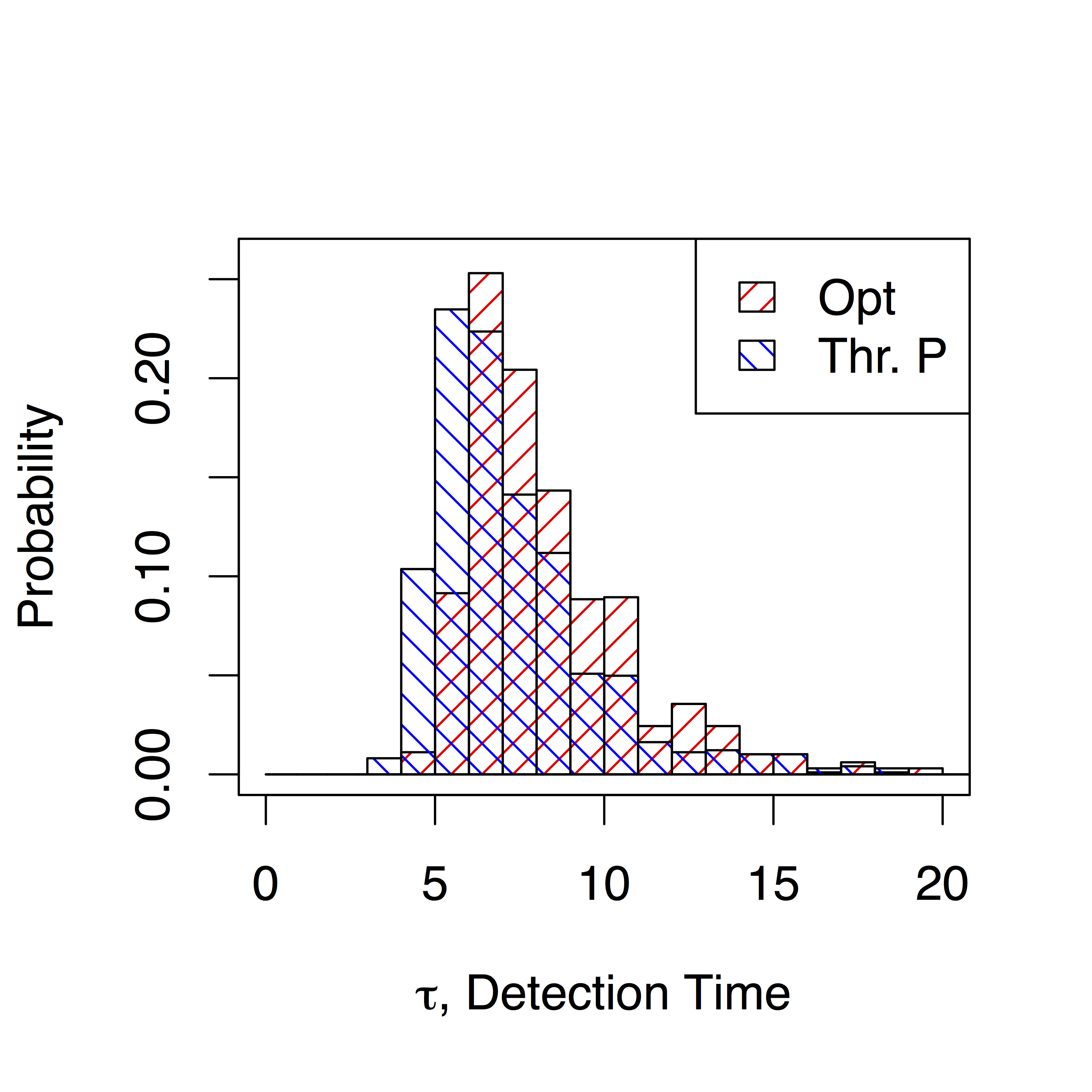}    
& \includegraphics[width=0.47\textwidth,trim=0.25in 0.25in 0.4in 0.2in]{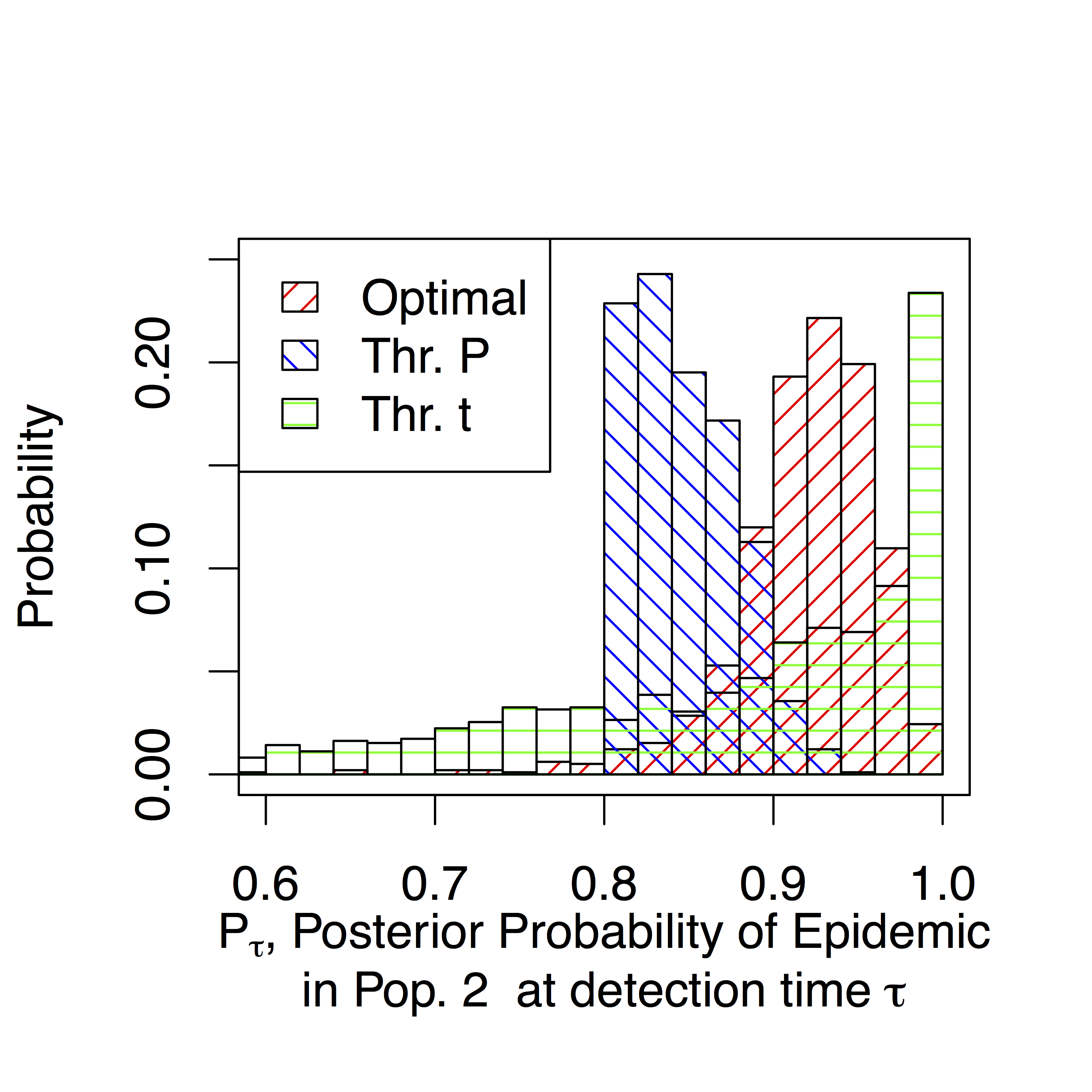}  \\   
\end{tabular}
\caption{Summary statistics of different detection strategies constructed from 1000 sample epidemic trajectories. The LP detection strategy is from Figure~\ref{fig: sample_detection_map}. Right: Distribution of detection times $\tau$;
Left: Distribution of posterior probability of outbreak in Pool 2 at detection time, $P_{\tau}$. \label{fig: hist_cost}}
\end{figure}

Returning to the full 3-D model with state $\mX$ we evaluate the resulting optimal detection strategy $\tau^*$  and proceed to compare its performance against the other potential detection rules discussed above. Specifically, the first two alternatives are a Threshold-P rule with $\bar{P}=0.8$ (declare an epidemic if its probability is above 80\%) and a Threshold-t strategy with $\bar{t} = 8$. The latter was found to be the best strategy among those that declare outbreak at a fixed stage. The last alternative is the LP strategy $\tau^{LP}$ from last section. Recall that $\tau^{LP}$ makes decisions while ignoring $S^{(1)}$. In that sense, when applied to the full 3-D model, it gives a simplified, but still adaptive, detection rule. To recap, Threshold-t strategy is completely non-adaptive; Threshold-P only relies on $P_t$; LP relies on $\{ I^{(1)}_t, P_t\}$, and Optimal strategy uses all of $\{ S^{(1)}_t, I^{(1)}_t, P_t\}$. 

To compare the performance of the above competing strategies, we fixed the initial condition at $S^{(1)}_0 = 1990, I_0^{(1)}=10$ and $P_0=0.1$, so that there are 10 infecteds in Pool 1 and 10\% prior probability of epidemic already in Pool 2. Then we simulated 1000 epidemic trajectories $\{\mx^n_{0:\tau} \}$, $n=1,\ldots, 1000$, emanating from this fixed initial condition up to the detection time $\tau$ (which depends in turn on the strategy used). Table \ref{table: comparison_strategies} then  presents the resulting summary statistics based on these frozen 1000 trajectories (note that there are no analytic formulas to obtain these metrics, so we have to resort to simulation).

The comparison is done in terms of several different metrics, including realized detection costs $c(\mathfrak{X}_{0:\tau^{(t)}})$, distribution of detection times $\tau$, and frequency of false alarms, represented by $d(\mathfrak{X}_\tau) = 1-P_\tau$ in our setup.
As expected, the Optimal strategy with detection time $\tau^*$ that directly optimizes the cost-benefit in the full model performs best. The corresponding expected costs are $V(\mx_0) \simeq 6.53$, with average detection time $E[\tau^*] \simeq 8.86$. It outperforms the Threshold-P strategy by about 7\% in terms of reducing detection costs, and the Threshold-t strategy by about 9\%. These are nontrivial cost savings which highlight the benefit of information fusion. Table~\ref{table: comparison_strategies} also shows that the 2-D LP approximation performs well in this example, generating very similar expected costs. At least for this case study, detection happens early enough that the branching process approximation of the outbreak works fine.

Recall that our model is stochastic and generates adaptive detection strategy. Hence the detection time $\tau^*$ is a random variable. As shown in Table \ref{table: comparison_strategies}, the corresponding standard deviation $StDev(\tau^*) \simeq 2.6$ is substantial. This illustrates the sub-optimality of the Threshold-t strategy that stops at a fixed $\bar{t}$ with $StDev(\bar{t}) = 0$ trivially. Not surprisingly, the ability to delay or speed up outbreak announcements based on latest data are crucial for optimizing policy making. We also note that compared to the Threshold-P strategy, the Optimal strategy tends to announce later, $E[ \tau^* ] \simeq 8.86 > 7.88 \simeq E[ \tau^{Thr-P}]$, this is also confirmed by the respective histograms of $\tau^{*}$ and $\tau^{Thr-P}$ in Figure~\ref{fig: hist_cost}. However, we emphasize that the detection rules do not have a clear ordering. In other words, the random variables $\tau^{*}$, $\tau^{Thr-P}$, etc., cannot be directly compared.

 A complementary metric of detection quality is provided by the probability of false alarms, $PFA := E[1-P_\tau]$. For the optimal strategy we find that $PFA^{*} = 8.2\%$. In contrast, for Threshold-P strategy, we have $PFA^{Thr-P} = 15.3\%$. Note that because we use a discrete-time model, at time of detection $P_\tau$ will strictly exceed the threshold $\bar{P} = 0.8$, hence $PFA^{Thr-P} < 1-\bar{P}$.
The histograms of $P_\tau$ are shown in Figure~\ref{fig: hist_cost} and confirm the qualitative difference among the detection strategies. The Threshold-P strategy only stops once $P_t > \bar{P}$, so that $P_{\tau}$ has support on roughly $[0.8, 0.9]$. In contrast, the adaptive Optimal (and LP) strategies, have a much wider range for $P_\tau$. In particular, sometimes epidemics are announced even before $P_t$ hits the level 0.8. \medskip

\begin{figure}[ht]
\centering
\includegraphics[width=0.58\textwidth,trim=0.3in 0.3in 0.3in 0.3in]{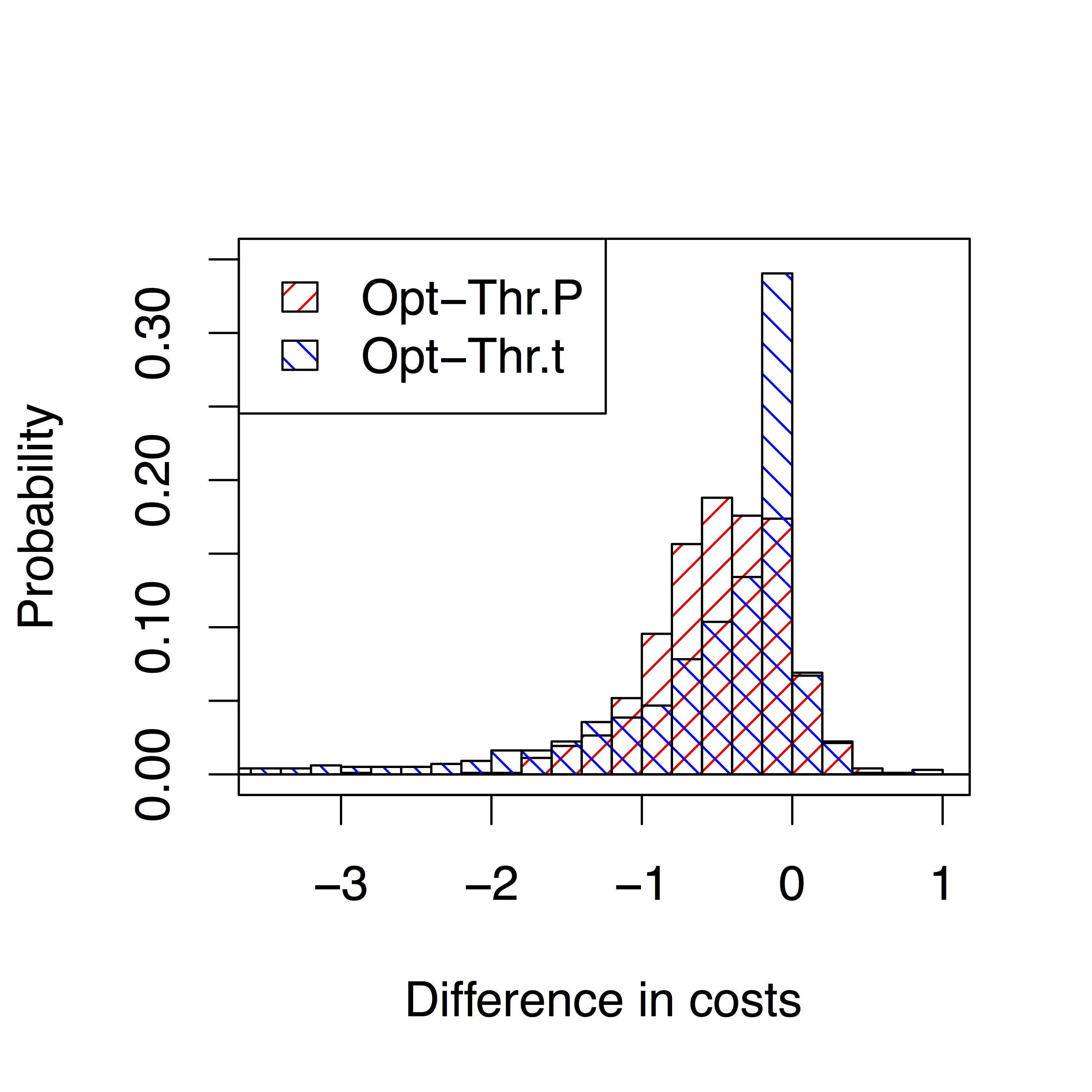}    
\caption{Relative realized detection costs across different strategies. The histogram shows the distribution of the difference in costs along the 1000 simulated trajectories, namely $c(\mx_{0:\tau^{*}})-c(\mx_{0:\tau^{Thr-P}})$, and $c(\mx_{0:\tau^*})-c(\mx_{0:\tau^{Thr-t}})$. \label{fig:cost-difference}}
\end{figure}

To further quantify the improvement provided by the Optimal detection rule, Figure \ref{fig:cost-difference} gives a scenario-by-scenario comparison of relative realized detection costs. Note that in hindsight, $\tau^*$ may sometimes perform worse that $\tau^{Thr-P}$ or even $\tau^{Thr-t}$. Figure~\ref{fig: hist_cost} plots the histogram of the difference in costs for each trajectory $\mx_{0:t}^n$, $n=1,\ldots,1000$, namely
$c(\mx_{0:\tau^{*}})$, $c(\mx_{0:\tau^{Thr-P}})$, and $c(\mx_{0:\tau^{Thr-t}})$. We find that the costs computed with Optimal/LP strategies are smaller than costs computed with Threshold strategies for more than 80\% of the trajectories.

To sum up, we observe material improvement from using Optimal detection rule in this case study. Moreover, the obtained detection rule is substantially different from the thresholding protocol. On the one hand, the adaptive detection time $\tau^*$ exhibits a wide spread and is highly non-constant across trajectories. On the other hand, the posterior probability of false alarms $P_{\tau^*}$ is also strongly variable. As a result, the average frequency of false alarms is drastically lowered relative to Threshold-P strategy, reducing overall expected costs.

%
%
%
%

\subsection{Effect of Detection Cost Parameters}
\begin{figure}[ht]
\centering
 \includegraphics[width=0.75\textwidth]{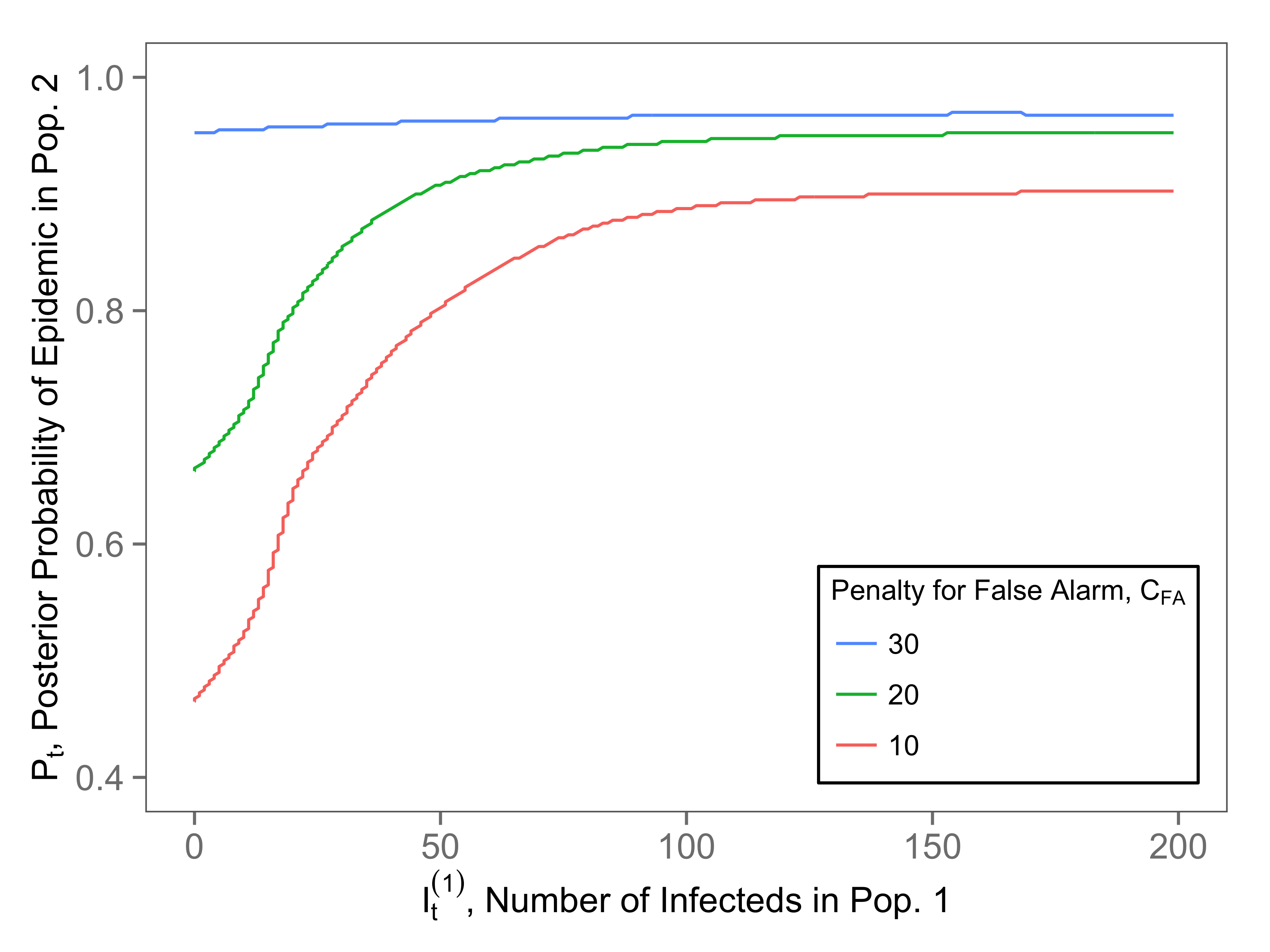}
\caption{Boundaries of detection maps $\partial \mathfrak{S}^{LP}_{20}$ constructed based on different penalties for false alarm, $C_{\text{FA}}$.}
\label{fig: costs_FA}
\end{figure}

\begin{table}[ht]
\centering
 \begin{tabular}{r|cc|cc|l}
\hline
\multirow{2}{*}{$C_{\text{FA}}$} 			&\multicolumn{2}{|c|}{$\tau^*$}	& \multicolumn{2}{|c|}{Cost} 	& \multirow{2}{*}{$PFA = E[1-P_{\tau^*}]$}   \\ 
			& Mean 	& StDev.		& Mean 	& StDev.				&   \\ \hline \hline
10  		& 6.84  	& 1.62   		& 5.32 	& 0.99				& 21.4\% \\ 
20		&  8.87 	& 2.60  &  6.54 & 1.71 & 8.3\%\\ 
30	   	& 9.61  & 2.79   & 7.21 & 2.22 &5.3\% \\ \hline

\end{tabular}
\caption{Summary statistics of the Optimal detection strategy  $\tau^*$ for different false alarm penalties $C_{\text{FA}}$.}
\label{table: comparison_costs}
\end{table}

The main parameter in our quickest detection setup is the ratio of the cost of false alarms and the cost of detection delay, $C_{\text{FA}}/C_{\text{Delay}}$. A high ratio penalizes premature announcements and requires more care in the assessment of the potential outbreak in Pool 2. A low ratio invites more aggressive actions. To better understand the role of this ratio, in Figure~\ref{fig: costs_FA} we show several detection boundaries $\partial \mS^{LP}$ corresponding to varying $C_{\text{FA}}$, while $C_{\text{Delay}}=1$ is kept fixed. As expected, a lower $C_{\text{FA}}$ enlarges the Announce set $\mS$. In particular, the boundary $\partial \mS$ shifts down and to the right. As a result, starting from a fixed location $(I^{(1)}_0, P_0)$, the stopping set $\mS$ will be reached sooner, so that $\tau$ decreases (in the sense of stochastic dominance for the corresponding random variables). This is confirmed in Table \ref{table: comparison_costs} that reports statistics for $\tau^*$ and various $C_{\text{FA}}$. We find that
 $E[\tau^*] = 8.86$ when $C_{\text{FA}} = 20$, but is only $E[\tau^*] = 6.84$ for $C_{\text{FA}} = 10$.  Simultaneously,
the frequency of premature announcements $PFA$ will increase. The precise relationship is however nonlinear. Lowering $C_{\text{FA}}$ from 20 to 10, the PFA rises dramatically to about 21\% from 8\%. Conversely, raising $C_{\text{FA}}$ to 30 only reduces PFA to 5.3\%. A common approach in the decision literature is to select a priori a desired level of PFA (say $PFA = 10\%$) and then numerically solve the inverse problem to obtain the corresponding $C_{\text{FA}}$ and hence the corresponding detection rule $\mS$.

%
%
%
%

\section{Numerical Implementation}
\label{section SRMC}

To find the detection maps  $\hat{\mathfrak{S}}_t$ for $t=1,2,\ldots,$ defined recursively in \eqref{eq: stoppingregion2} we use approximate dynamic programming techniques. In particular, we rely on the Regression Monte Carlo approach \cite{Egloff05,GL14} to approximate the conditional expectation map over $\mx \in \cX$.

\subsection{Regression Monte Carlo}
\label{sec: RMC}

For the remainder of this section the auxiliary ``time'' variable $t$ is fixed and the goal is to approximate the conditional expectation  $q(t,\mx) := E[ c(\mathfrak{X}_{0:\tau^{(t)}} | \mathfrak{X}_0 = \mx]$ in \eqref{eq:c-tau-star}.  Recall that at step $t$, detection rules are restricted to satisfy $\tau^{(t)} \le t$.
The Regression Monte Carlo technique approximates $q(t,\cdot)$ by a predicted surrogate value $\hat{q}(t,\cdot)$  which is based on a statistical regression framework.

The surrogate prediction is built using data simulated from the specified model. To do so, a design $\cZ := \{\mx_0^n,\ n=1,\ldots,N\}$ of $N$ locations is first generated. Next, we generate the corresponding scenarios $\{\mathfrak{X}^n_{0:t}\}$ with the initial value $\mathfrak{X}^n_0= \mx_0^n$, one scenario for each initial location. Define
\begin{equation} \label{eq: def.tau_m}
\tau^n_{t} := \min \{s \ge 1: \mathfrak{X}^n_{s} \in \mathfrak{S}_{t-s}\},
\end{equation}
which leads to path-wise waiting costs $q^n := c(\mathfrak{X}^n_{0:\tau^n_t})$ using formula \eqref{eq: costs} on the $n$-th scenario. The aggregate dataset is
\begin{equation} \label{eq: design}
Z=\left\{ \left( \mx_0^n, q^n \right), n=1,\ldots,N\right\}.
\end{equation}

The construction of $\hat{q}(t,\cdot)$ then involves response surface modeling, i.e.~determining the relationship between the initial condition $\mx$ and the mean of the sampled  $Q| \mx \equiv c(\mathfrak{X}_{0:\tau_t})$. Statistically, we start with
\begin{align}
{Q}|\mx =q(t,\mx)+\epsilon,
\label{eq:regression_model}
\end{align}
where $q(t,\cdot)$ is the true response surface, $Q = c(\mathfrak{X}_{0:\tau^{(t)}})$ are random scenario-based costs, and $\epsilon$ are mean-zero residuals with variance $\sigma^2$ arising from Monte Carlo simulations. Empirically, \eqref{eq:regression_model} translates into regressing $\{q^n\}$ on $\{\mx_0^n\}$, $n=1, \ldots, N$; this step is discussed in section \ref{sec: LocalRM}.
%
%
After determining $\hat{q}$, and using \eqref{eq: stoppingregion2} the estimated detection rule $\hat{\mathfrak{S}}_t$  is
\begin{equation}
\hat{\mathfrak{S}}_t:=\left\{\mx:  \hat{q}(t,\mx) - d(\mx) > 0\right\}.
\label{eq: stoppingregion_RMC}
\end{equation}

\begin{figure}[ht]
\centering
\includegraphics[width =0.75 \textwidth]{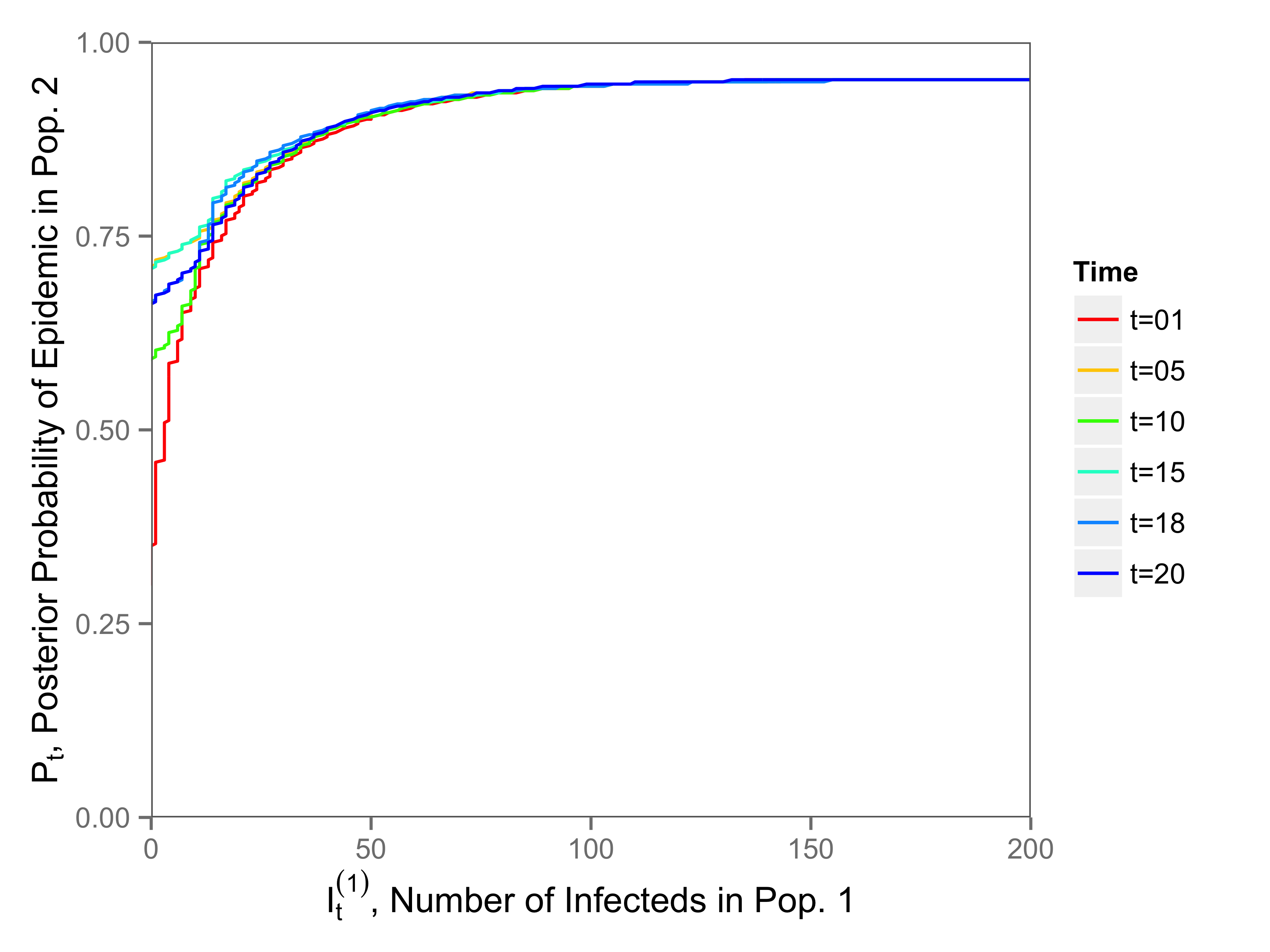}
\caption{Convergence of the detection boundaries $\partial \hmS_t$ over $t=1$ to $t=20$ for the 2-D LP detection rule from Section \ref{sec:case-study}.
\label{fig: sample_detection_map_t_improv}}
\end{figure}

The above provides a recipe to obtain an (approximate) $\hmS_{t}$ using the collection of detection rules $\hmS_{1:t-1}$. Iterating over $t$, yields the sequence of detection maps $\hmS_{t}$ for $t=1,2,\ldots$. 
We recall that as $t\to \infty$, we expect $\hmS_t$ to stabilize and tend to a time-invariant detection map. Such convergence is illustrated in Figure~\ref{fig: sample_detection_map_t_improv}, where we trace the boundaries $\partial \hmS_t$ for $t=1,\ldots,20$. Convergence takes hold after about 15 iterations and suggests that $\hmS_{20} \simeq \mS$; this is what we used for Figures \ref{fig: sample_detection_map}-\ref{fig: costs_FA} where the boundary of $\hmS_{20}$ was taken as the final output of the Algorithm.

The detection rule \eqref{eq: def.tau_m} is time-dependent since it utilizes a new  $\hmS_{t-s}$ at each stage $s$. Model predictive control simplifies this feature with a time-invariant rule that simply utilizes $\hmS_{t-1}$ (that we relabel as $\hmS^{(t-1)}$ for typographical distinction). Indeed, as Figure \ref{fig: sample_detection_map_t_improv} shows, the early maps $\hmS_1, \hmS_2, \ldots$, are not as accurate as $\hmS_{t-1}$ for $t$ large, so it makes sense to completely ``forget'' them and rely just on the last iteration step. Accordingly, we implement a blend of \eqref{eq:mpc} and \eqref{eq: value_function3} by first using \eqref{eq: def.tau_m} over $t=1,2,\ldots, t^*$ and then switching to a receding-horizon rule
\begin{equation} \label{eq: def.tau_mpc}
\tau^{MPC}_{t} := \min \{s \ge 1: \mathfrak{X}_{s} \in \mathfrak{S}_{t-1}\}, \qquad t=t^*,t^*+1, \ldots.
\end{equation}
The above MPC iterations are terminated once $\hat{q}(t,\mx)$ and $\hat{q}(t+1,\mx)$ do not change much, namely $\| \hat{q}(t,\cdot) -\hat{q}(t+1,\cdot) \|_{L^\infty} < Tol$ for a specified tolerance level $Tol$.

\subsection{Regression Model}
\label{sec: LocalRM}

Because we have limited a priori knowledge about the structure of the detection rule, it is preferable to work with a nonparametric regression architecture for $q(t,\mx)$. (For example a linear regression model for $q$ would imply that $\mS$ in \eqref{eq: stoppingregion_RMC} is defined through linear constraints, i.e.~forms a simplex in $\cX$.) In addition, nonparametric regression is typically more robust for dealing with the non-Gaussian residuals $\epsilon$ that arise in our model.

There are numerous nonparametric regression frameworks that can be used, including splines, Gaussian processes, or generalized additive models; see e.g.~the classic monograph \cite{FHT}. Note that even though $\mx \mapsto V(\mx)$ is continuous, some discontinuous response surfaces might also be helpful, such as random forests or dynamic trees \cite{GL14}. In the present article we take up a variant of local linear regression, known as Loess. Loess fits weighted linear regression models to localized subsets of data, determined using a kernel function, specifically a $k$-nearest-neighbor algorithm \cite{1988}. Compared to classical linear models, Loess better handles outliers and heteroscedasticity, and also does not make assumptions about the global shape of the response surface.

The Loess response model is of the form \begin{align}
\label{eq:loess}\hat{q}(t,\mx)= \sum_{i=1}^r \hat{\beta}_i(\mx) B_i(\mx)
\end{align}
 where $B_i(\cdot)$ is the set of $r$ pre-specified basis functions and $\hat{\beta}_i$ are estimated regression coefficients at $\mx$.
 Given input matrix $\vec{X}$ and matching response vector $Q$, $\hat{\mathbf{\beta}}$ is  fitted
using local least-squares minimization
\begin{equation}
\hat{\mathbf{\beta}}(\mx) := \arg \min_{\vec\beta \in \mathbb{R}^r}  K_{\lambda}(\mx, \vec{X}) (Q-B(\vec{X})^T \vec{\beta} )^2,
\end{equation}
where $K_{\lambda}(\mx, \vec{X})$ is the weighting kernel. The idea behind the kernel is to base the predicted $\hat{q}(t,\mx)$ on the samples in the neighborhood of $\mx$, weighted by their distance from $\mx$ \cite[Sec.~2.8.2]{FHT}. The size of the neighborhood is controlled by the smoothing parameter $\lambda$. If $\lambda<1$, only a proportion $\lambda$ of the samples will be used in fitting. The smaller $\lambda$, the more ``wiggly'' the fit $\hat{q}(t,\cdot)$ is going to be since fewer samples are used in computing $\hat\beta(\mx)$.
 Loess can be viewed as a special kernel regression method, with the prediction being a weighted average of the responses $q^n$:
 $\hat{q}(t,\mx) = \sum_n l_n(\mx) q^n$
for the equivalent kernel $l(\cdot)$. In our numerical examples, we use the implementation of Loess provided in the \texttt{R} by the built-in package \texttt{stats} \cite{R}, which uses a tri-cubic kernel and linear, first-order basis functions; the smoothing parameter was $\lambda= 0.4$.

\subsection{Experimental Design}
\label{sec: SRMC}

The aim of the response surface is to maximize the accuracy of $\hmS_t$. This is equivalent to maximizing model fidelity along the boundary of the detection map. Statistically, for a localized response surface, accuracy is primarily driven by the local density of the input data that is specified by the experimental design $\cZ$. Hence, to maximize our confidence regarding the boundary of ${\mathfrak{S}}_t$ in \eqref{eq: stoppingregion_RMC},  we generate appropriate, adaptively chosen experimental designs $\cZ$. This is achieved using the Sequential RMC framework introduced by \cite{GL14}. SRMC uses tools from active learning/Bayesian optimization to gradually \emph{grow} the design $\cZ$ so as to zoom-in to the boundary of $\hmS_t$. This is done by first quantifying the accuracy of the existing response surface, and then adding new design sites so as to maximize information gain. See \cite{GL14,HL15} for details. The SRMC approach is illustrated in Figure \ref{fig: sample_detection_map} where the adaptively generated experimental design $\cZ$ (of size 2000 in the figure) is highly concentrated around the detection boundary $\partial \mS$. This targeted sampling of outbreak scenarios allows for more efficient estimation, in particular lowering the local standard errors $\hat{v}(\mx)$ along $\partial \hmS_t$, cf.~the right panel of Figure \ref{fig: sample_detection_map}.

In \eqref{eq: stoppingregion_RMC} the boundary of $\hmS_t$ corresponds to the regions of $\cX$ where the cost difference between immediate detection and waiting is zero. Hence, we aim to have more design points in regions where $\{ \hat{q}(t,\mx) - d(\mx) \simeq 0\}$. To this end, we define the ``posterior'' measure of response surface accuracy via
\begin{equation}
p(\mx) := \Phi \left( \frac{ -|\hat{q}(t,\mx) - d(\mx)|}{\sqrt{ \hat{v}(\mx)}} \right),
\label{eq: normal_approx}
\end{equation}
where $\Phi$ is the standard normal cdf and the predictive variance $\hat{v}$ measures the standard error of the surrogate prediction,
\begin{align}
\hat{v}(\mx)= \hat{\sigma}^2(\mx) \| l(\mx) \|^2, \label{eq: variance_fitted}
\end{align}
with $\hat{\sigma}^2(\mx)$ the estimated variance of $\epsilon$ around $\mx$ in \eqref{eq:regression_model}, see  \cite[Sec 6.1.2]{FHT}.

The motivation for \eqref{eq: normal_approx} is that $p(\mx)$ mimics the Bayesian posterior probability of estimating the wrong \emph{sign} (conditional on the samples in $\cZ$) of $q(t,\mathbf{\mx}) - d(\mathbf{\mx})$, assuming that the posterior distribution is Gaussian with the empirical mean $\hat{q}(t,\mx)$ and variance $\hat{v}(\mx)$.

The defined metric $p(\cdot)$ serves as a guide to augment new design locations. Namely, it defines an acquisition function $w(\mx)$ for greedily growing $\cZ$, similar to active learning methods \cite{Mackay92}. The acquisition function is highest in the regions where $p(\mx)$ is close to $0.5$ which correspond to $\partial\hat{\mathfrak{S}}_t$. Our main choice is
\begin{align}
w^{\text{min}}( \mx)=&\min\left[p( \mx), 1- p( \mx)\right]. \label{eq: min}
\end{align}
Alternatives include the Gini weights $w^{\text{gini}}(\mx)=p( \mx) \left(1-p( \mx)\right) $ 
and Entropic weights $w^{\text{Ent}}(\mx)=-p( \mx) \log p(\mx) - (1-p( \mx)) \log(1-p( \mx))$.

To speed up the response surface modeling, which requires refitting of $\hat{q}(t,\cdot)$ multiple times, we used batch steps, incrementally working with designs $\cZ^{(N)}$ of size $N=N_0, N_0+N',\ldots, N^{end}$. At each sequential design iteration, an additional $N'$ design points $\{\mx_0^n\}_{n=N+1}^{N+N'}$ are added to existing $\cZ^{(N)}$. Those are sampled multinomially in proportion to the acquisition function $w(\cdot)$ from a candidate set ${X}_{\text{finite}}$. Both the initial design $\cZ^{(N_0)}$ and the candidate sets $X_{\text{finite}}$ are generated using Latin hypercube sampling (LHS) of size $D$ from $\cX$.  The overall procedure, summarized in Algorithm~\ref{algorithm: SRMC}, finally refits at each iteration the Loess model for $\hat{q}$ (and hence $\mathfrak{S}_t$), grows the experimental design $\cZ^{(N+N')} = \cZ^{(N)} \cup \{\mx_0^n\}_{n=N+1}^{N+N'} $ and recomputes the acquisition function \eqref{eq: min}. As the design size gets larger, we expect that the implied empirical estimate $\partial\hat{\mathfrak{S}}^{(N)}_t$ gets closer to the true $\partial\mathfrak{S}_t$.

\begin{remark}
One can apply standard, non-sequential RMC by skipping the inner while loop (steps 7-15) in Algorithm \ref{algorithm: SRMC}. This reduces to building a response model on a pre-specified (possibly randomized) design $\cZ := \{ \mx_0^n\}_{n=1}^{N_0}$, keeping all other steps as is.
\end{remark}

For the detection map in Figure \ref{fig: sample_detection_map} in Section \ref{sec:case-study} we used an initial design of $N_0 = 200$, which was grown over 10 iterations with $N'=200$ to a final design of $N^{end} = 2000$. The acquisition function was $w^{\text{min}}$ and the candidate sets $X_{\text{finite}}$ of size $D=2500$ were generated with LHS. Since detection happens while $I^{(1)}$ is still relatively small, we restricted the response surface regression domain to $I^{(1)} \in \{0,1,\ldots,400\}, S^{(1)} \in \{1000,\ldots,2000\}$. Lastly we note that the  method is still computationally intensive, with the bulk of the effort spent on generating $T \cdot N^{end}$ scenarios of $\mX$;  running times (on a 8-core 2.27GHz machine with 12GB of RAM) were about 20 minutes.

\section{Discussion}
\label{section: discussion}

We have presented a framework for optimal detection of epidemics in a coupled meta-population model. Our approach explicitly takes into account cost-benefit considerations regarding announcement of an epidemic, as well as spatial dependence across susceptible pools. Given the information about two populations and characteristics of the infection, our algorithm produces the full detection map which can then be used repeatedly. We demonstrate that information about the epidemic in one pool can be used to lower the detection costs in another pool, realizing savings compared to traditional threshold-type detection methods.

Since our dynamic optimization approach is entirely simulation-based it is unusually flexible. Indeed, the precise underlying epidemic model of $\mX$ is not crucial, since Algorithm \ref{algorithm: SRMC} only requires the ability to generate its trajectories. In fact,  the computational complexity of our algorithms is tied not to the dynamics of $\mX$ but to its dimensionality. In Section \ref{sec:case-study}, we had $\dimn( \mX) = 3$; based on our experience with RMC in \cite{LN10,GL14}, the present approach can straightforwardly handle up to 6-8 dimensions. In high dimensions, extra care must be applied for generating the experimental designs $\cZ$ since the concept of neighborhoods underlying nonparametric local regression breaks down. For example, Loess regression performs poorly if the dimension of the data is higher than 4-5.

The presented SIR framework gives a basic mechanistic description of disease progression that is obviously not very realistic. More sophisticated versions might allow for further compartments (such as Exposed or Diseased individuals), age stratification,  and heterogeneity among the meta-populations. One could also include further transitions beyond \eqref{eqn: reaction_channels}, such as immigration $\emptyset \to I^{(k)}$, immunity lapses $R^{(k)} \to S^{(k)}$, or vaccination $S^{(k)} \to R^{(k)}$. Introducing immigration would allow for endogenous epidemic in Pool 2, removing the assumption that outbreaks always start in Pool 1 and then spread to Pool 2. The constant transition rates used can be replaced with seasonal patterns, stochastic fluctuations \cite{IonidesKingMeasles10}, or hierarchical Markov structures \cite{lin2013sequential}.

 Alternatively, one can also imagine more sophisticated models for the outbreak pseudo-posterior  $P_t$ -- recall that the proposed one was largely for convenience than any realism. For example, the Gaussian noise $\delta_t$ in the dynamics of $P_t$ that was used in the case study may be better modeled via a Beta distribution (which arises naturally as conjugate to the Poisson increments of the fully-observed stochastic SIR model \cite{lin2013sequential}). Overall, the key requirement is the Markov structure which makes it possible to use regression against $\mX$ to describe the detection rule. The Markovianity requirement can be partly relaxed if one is willing to accept approximately-optimal solutions. Indeed, one can always project the optimal detection rule into the smaller space of rules that only depend on some subset $\mX'$; in other words restricting the detection map to only take into account some of the state-space dimensions. This idea was already discussed in Section \ref{sec:case-study} where we described the sub-optimal LP strategy.


Second, one may modify the cost structures \eqref{eq: cost_today}-\eqref{eq: costs} to better capture the desired detection goals. The presented costs were motivated by their classical analogues in sequential change-point detection, but might not be the most appropriate for public health contexts. For example, there is some leeway in what constitutes an outbreak. In \eqref{eq: costs}, the threshold was zero, i.e.~even a single infected individual in Pool 2 was reportable. One can use thresholds $\bar{I}$ other than zero, so that an outbreak is reportable only when  $I_t^{(2)} > \bar{I}$, otherwise an announcement is treated as premature. Similarly, the waiting cost in \eqref{eq: costs} was constant; it may be more realistic to make it proportional to $I_t^{(2)}$, which would correspond to fixed costs per infected.


For such  more general formulations, the costs $d(\mX)$ and $c(\mX)$ would no longer be  functions of $P_t$, and one would need to work with the full posterior distribution $\pi_t$ of $I^{(2)}_t | \cG_t$. The RMC framework could still be usable, namely we may use particle filtering \cite{LudkovskiOspi} to obtain $\pi_t$ along a simulated trajectory of the underlying epidemic model. Certainly, particle filtering can become computationally expensive, making efficient inference essential. We refer to \cite{Niemi2014,lin2013sequential} for some recent implementation strategies in this direction that specifically target epidemic models. The integrated sequential inference + optimization model would then allow to treat a partially observed version of a $K$-pool SIR model of \eqref{eqn: reaction_channels}, and ultimately a larger-scale setup such as influenza surveillance across all 50 states, cf.~Figure \ref{fig:epidemic-movie}.




\section*{Acknowledgement}
This material is based upon work supported by the National Science Foundation under Grant No. ATD - 1222262.


\bibliographystyle{elsarticle-num}
\bibliography{5_references}

\newpage
\appendix
\section{Algorithms}
\label{appendix}

\renewcommand{\thealgorithm}{A.\arabic{algorithm}}

\begin{algorithm*}[!ht]
\caption{Path and Cost Generation}
\begin{algorithmic}[1]
\Require $\{\mx_0^n\}_{n=1}^N$, $\mathfrak{S}_{0:t-1}$
\For {$n =1,\ldots, N$}
\State $s \gets 1$
\While {$s \le t$}
\State Simulate the next state $\mx_s^n \sim p_1(\cdot| \mx_{s-1}^n)$
\If {$\mx_s^n \in \mS_{t-s}$}  Break \EndIf
\State $s \gets s+1$
\EndWhile
\State $\tau^n_t \gets s$
\State Compute $q^n \equiv c(\mx^n_{0:\tau^n_t})$ using formula \eqref{eq: costs}
\EndFor \\
\Return  $\{ \left(\mx^n_0, q^n \right) \}_{n=1}^N$
\end{algorithmic}
\label{algorithm: cost_gen}
\end{algorithm*}


\begin{algorithm*}[!ht]
\caption{Sequential Regression Monte Carlo}
\label{algorithm: SRMC}
\begin{algorithmic}[1]
\Require $C_{\text{FA}}, C_{\text{Delay}}$, $N_0$, $N'$, $N^{end}$, D
\State $\hmS_0 \gets \cX$
\For {$t=1,2,\ldots$}
\State Generate experimental design $\{\mx_0^n,\ n=1,\ldots,N_0\}$
\State Compute scenario costs $q^n = c(\mathfrak{X}^n_{0:\tau^n_t})$ for $n=1,\ldots,N_0$ using Algorithm~\ref{algorithm: cost_gen} and $\hat{\mathfrak{S}}_{0:t-1}$
\State $Z\gets \left\{ \left( \mx_0^n, q^n\right)\right\}_{n=1}^{N_0}$
\State Regress $\{q^n\}$ on $\{\mx_0^n\}, \ n=1, \ldots, N_0$ using Loess \eqref{eq:loess}
\State Initialize $N \gets N_0$
\While {$N < N^{end}$}
\State Generate $X_{\text{finite}}$ of size $D$ using Latin Hypercube Sampling on $\cX$
\State Compute the acquisition weights $w(\mx) \ \forall \mx \in X_{\text{finite}}$ via \eqref{eq: min} and \eqref{eq: normal_approx}
\State Sample $\{\mx_0^n\}_{n=N+1}^{N+N'}$ from $X_{\text{finite}}$ using weights  $w(\mx)$
\State Simulate the costs $q^n$, $n=N+1, \ldots, N+N'$ using Algorithm~\ref{algorithm: cost_gen}
\State $Z \gets Z \cup \left\{ \left( \mx_0^n, q^n\right)\right\}_{n=N+1}^{N + N'}$
\State Update the Loess regression model \eqref{eq:loess} using the latest $Z$
\State  $N\gets N+N'$
\EndWhile
\State $\hat{\mathfrak{S}}_t \gets \{ \mx \in \cX : \hat{q}(t,\mx) - d(\mx) > 0 \}$, cf.~\eqref{eq: stoppingregion_RMC}
\EndFor
\end{algorithmic}
\end{algorithm*}


\end{document}